\documentclass[aps,prb,preprint,superscriptaddress]{revtex4}

\usepackage{amsmath,amssymb}
\usepackage{graphicx,color}
\usepackage{dcolumn}
\usepackage{bm}

\def\e{\mathrm{e}}
\def\i{\mathrm{i}}

\def\nn{\nonumber}

\def\sgn{\mathrm{sgn}}

\begin{document}


\title{Universal electric current 
of interacting resonant-level models 
with asymmetric interactions: 
An extension of the Landauer formula}

\author{Akinori Nishino}\email{nishino@kanagawa-u.ac.jp}
\affiliation{Faculty of Engineering, Kanagawa University, 
3-27-1, Rokkakubashi, Kanagawa-ku, 
Yokohama-shi, Kanagawa, 221-8686, Japan}

\author{Naomichi Hatano}
\affiliation{Institute of Industrial Science, 
The University of Tokyo, 
4-6-1 Komaba, Meguro-ku, Tokyo, 153-8505, Japan}
\affiliation{Department of Physics, Washington University, 
St.~Louis, Missouri, 63130, USA}

\author{Gonzalo Ordonez}
\affiliation{Department of Physics and Astronomy,
Butler University,
4600 Sunset Ave., Indianapolis, Indiana 46208, USA}

\date{\today}

\begin{abstract}
We study the electron transport in open quantum-dot systems
described by the interacting resonant-level models
with Coulomb interactions.
We consider the situation in which the quantum dot is
connected to the left and right leads asymmetrically.
We exactly construct many-electron scattering eigenstates
for the two-lead system, 
where two-body bound states appear as a consequence of 
one-body resonances and the Coulomb interactions.
By using an extension of the Landauer formula, 
we calculate the average electric current 
for the system under bias voltages
in the first order of the interaction parameters.
Through a renormalization-group technique,
we arrive at the universal electric current, where
we observe the suppression of the electric current 
for large bias voltages, i.e., negative differential conductance.
We find that the suppressed electric current is restored 
by the asymmetry of the system parameters.

\end{abstract}

\maketitle


\section{Introduction}

In the last two decades, much progress has been made
in the experimental studies of the electron transport 
in nanoscale devices~\cite{%
GoldhaberGordon-Shtrikman-Mahalu-AbuschMagder-Meirav-Kastner_98Nature,%
Cronenwett-Oosterkamp-Kouwenhoven_98Science,%
Wiel-Franceschi-Fujisawa-Elzerman-Tarucha-Kouwenhoven_00Science,%
Kretinin-Shtrikman-GoldhaberGordon-Hanl-Weichselbaum-vonDelft-Costi-Mahalu_11PRB}.
In the systems smaller than the coherent length, quantum effects
are observed in the electron 
transport~\cite{Yacoby-Heiblum-Mahalu-Shtrikman_95PRL,%
Schuster-Buks-Heiblum-Mahalu-Umansky-Shtrikman_97Nature}.
In order to analyze it
beyond the linear-response regime theoretically, we need to treat 
nonequilibrium steady states realized in {\it open} quantum systems.
The Landauer formula~\cite{Datta,Imry} enables the calculation of 
the electric current flowing across nanoscale samples in noninteracting cases, 
which indicates that the nonequilibrium steady states are 
scattering states in the open quantum systems.
Indeed, the transport properties such as the electrical conductance and 
the electric-current noise are determined by the scattering matrix~\cite{%
Yurke-Kochanski_90PRB,Buttiker_92PRB,Blanter-Buttiker_00PR}.
To investigate interacting cases, the Keldysh formalism of 
the nonequilibrium Green's function has been developed~\cite{%
Caroli-Combescot-Nozieres-SaintJames_71JPC,%
Meir-Wingreen-Lee_91PRL,Meir-Wingreen_92PRL,%
Hershfield-Davies-Wilkins_92PRB,Haug-Jauho}.
It has provided a standard tool for the study of the Kondo effect 
measured as a conductance peak in semiconductor 
quantum dots (QDs)~\cite{Ng-Lee_88PRL,Wingreen-Meir_94PRB}.
On the other hand,
we have proposed an extension of the Landauer formula
to interacting cases~\cite{%
Nishino-Imamura-Hatano_09PRL,Nishino-Imamura-Hatano_11PRB},
and have shown that the scattering states are essential
in the interacting cases as well.

The interacting resonant-level model (IRLM) is one of the standard
testbeds for such studies of the open QD systems with interactions.
The original IRLM, which consists of a single impurity
coupled to a conduction band, was introduced for 
studying the Kondo problem in equilibrium 
systems~\cite{Filyov-Wiegmann_80PLA}.
Recently, the IRLM with {\it two} external leads has been employed 
as a minimal model of open QD systems with Coulomb interactions;
it plays an important role in verifying
the theoretical approaches such as 
the nonequilibrium Bethe-ansatz approach~\cite{Mehta-Andrei_06PRL}, 
the perturbation theory with the numerical renormalization
group~\cite{Borda-Vladar-Zawadowski_07PRB},
a new method called impurity conditions~\cite{Doyon_07PRL}
and the time-dependent density-matrix renormalization-group
method~\cite{Boulat-Saleur-Schmitteckert_08PRL}.

A remarkable feature of the two-lead IRLM 
is the appearance of {\it negative differential conductance}, that is, 
the suppression of the electric current 
due to the Coulomb interaction for large bias voltages.
Clearly, this is a phenomenon out of the linear-response regime.
To see the feature and to compare the results obtained by different approaches,
the {\it universal electric current} characterized by 
a single scaling parameter $T_{\rm K}$ is 
useful~\cite{Doyon_07PRL,Doyon-Andrei_06PRB,Golub_07PRB}.
Indeed, it is found that, for large bias voltages $V$, 
the universal electric current shows a power-law decay
$\langle I\rangle\propto (V/T_{\rm K})^{-U/\pi}$ 
with the parameter $U$ of the Coulomb interaction~\cite{%
Doyon_07PRL,Boulat-Saleur-Schmitteckert_08PRL}.

In the previous papers~\cite{%
Nishino-Imamura-Hatano_09PRL,Nishino-Imamura-Hatano_11PRB},
we proposed an extension of the Landauer formula
with many-electron scattering eigenstates.
We considered the two-lead IRLM with linearized dispersion relations
and gave {\it exact} many-electron scattering eigenstates in explicit forms. 
This is in contrast to the previous studies~\cite{%
Aharony-EntinWohlman-Imry_00PRB,Goorden-Buttiker_07PRL,%
Lebedev-Lesovik-Blatter_08PRL} of the scattering problems for other QD systems,
which include integrals or matrix inversions.
The explicit $N$-electron scattering eigenstates enabled us 
to calculate the quantum-mechanical expectation value of 
the electric current, which we called the $N$-electron current.
By taking the electron-reservoir limit $N\to\infty$ of the $N$-electron 
current, we obtained the average electric current 
for the system under finite bias voltages.
It is clear that the way of realizing the nonequilibrium steady states
in our extension of the Landauer formula is different from the Keldysh 
formalism~\cite{Caroli-Combescot-Nozieres-SaintJames_71JPC,%
Meir-Wingreen-Lee_91PRL,Meir-Wingreen_92PRL,%
Hershfield-Davies-Wilkins_92PRB,Haug-Jauho}.
By employing a renormalization-group technique with the Callan-Symanzik 
equation~\cite{Doyon_07PRL,Golub_07PRB}, 
we arrived at the universal 
electric current in the first order of the interaction parameter $U$. 
We found that
the negative differential conductance of the universal electric current 
is characterized by the same scaling parameter $T_{\rm K}$
as that obtained by other approaches~\cite{Doyon_07PRL,Golub_07PRB,%
Karrasch-Andergassen-Pletyukhov_10EPL}.
We remark that the apparent inconsistency in Ref.~\onlinecite{Nishino-Imamura-Hatano_09PRL}
pointed out in Ref.~\onlinecite{Andergassen-Pletyukhov-Schuricht-Schoeller-Borda_11PRB}
is removed in the level of the universal electric current~\cite{Schuricht_2012}.

In the present paper, we study the two-lead IRLM in which the QD is connected
to the two external leads {\it asymmetrically}.  
The effect of the asymmetry of the QD systems is observed in experiments.
For example, in semiconductor QDs~\cite{%
Wiel-Franceschi-Fujisawa-Elzerman-Tarucha-Kouwenhoven_00Science,%
Kretinin-Shtrikman-GoldhaberGordon-Hanl-Weichselbaum-vonDelft-Costi-Mahalu_11PRB}, 
the asymmetry of the lead-dot couplings
causes breaking of the unitary limit of the Kondo effect,
which is theoretically understood in the linear-response 
regime~\cite{Ng-Lee_88PRL,Wingreen-Meir_94PRB}.
In the present study, we investigate the effect of the asymmetry
on electron transport out of the linear-response regime.
One of the theoretical difficulties of the asymmetric cases is that the even-odd 
transformation, which maps the two-lead IRLM to two single-lead systems~\cite{%
Mehta-Andrei_06PRL,Nishino-Imamura-Hatano_09PRL,Nishino-Imamura-Hatano_11PRB}, 
does not work.
The application of the Bethe-ansatz approach~\cite{Mehta-Andrei_06PRL}
has been restricted to the cases in which the even-odd transformation works.
The construction of exact many-electron scattering eigenstates 
for such pure two-lead systems is established 
for the first time in this paper.

Through the extension of the Landauer formula 
and a renormalization-group technique,
we obtain the universal electric current for the asymmetric cases
in the first order of the Coulomb-interaction
parameters $U_{1}$ and $U_{2}$.
The universal electric current is characterized by 
the two renormalized parameters $T_{1}$ and $T_{2}$.
The sum $T_{\rm K}=T_{1}+T_{2}$ provides 
a scaling parameter for the bias voltage $V$,
which is similar to the symmetric cases.
Our universal electric current has the same functional form 
as that obtained by the renormalization-group approach~\cite{%
Karrasch-Andergassen-Pletyukhov_10EPL,%
Andergassen-Pletyukhov-Schuricht-Schoeller-Borda_11PRB},
although there is a difference 
in our calculation of the renormalized parameters $T_{1}$ and $T_{2}$.
As we will point at the end of Section~\ref{sec:electric-current},  
this leads to a critical difference in the predicted behavior 
of the universal electric current.
The suppressed electric current due to the Coulomb interaction
is restored by the asymmetry of the system parameters.
To clarify the relation between the asymmetry of the system parameters and
the restoration of the suppressed electric current,
we introduce the asymmetry parameter 
$\delta=(U_{1}-U_{2})(T_{1}-T_{2})/(2T_{\rm K})$ taking the value 
in the range $0\leq \delta<\Bar{U}$
with the average interaction $\Bar{U}=(U_{1}+U_{2})/2$.
In fact, in the first order of $U_{1}$ and $U_{2}$,
the power-law decay of the universal electric current is given
by $\langle I\rangle\propto (V/T_{\rm K})^{-(\Bar{U}-\delta)/\pi}$,
which indicates that the restoration of the suppressed electric current
occurs with both asymmetric Coulomb interactions
and asymmetric lead-dot couplings.
The restoration was reported to happen even for symmetric lead-dot couplings 
in Refs.~\onlinecite{%
Karrasch-Andergassen-Pletyukhov_10EPL} and
\onlinecite{Andergassen-Pletyukhov-Schuricht-Schoeller-Borda_11PRB},
but we presume that this was due to higher orders of 
the interaction $U_{\ell}$ in the renormalized parameters $T_{1}$ and $T_{2}$.

The exact many-electron scattering eigenstates tell us much 
about the transport properties of interacting electrons in the open QD systems.
We notice that the scattering processes in which the set of wave numbers
of incident plane waves is not conserved are essential in interacting cases.
The explicit form of the scattering eigenstates indicates that, 
due to the Coulomb interactions, the incident plane-wave states are 
partially scattered to {\it many-body bound states} that decay 
exponentially as the electrons separate from each other.
Indeed, the two-body bound states appear in the two-lead IRLM;
each term of the $N$-electron scattering states
is characterized by the configuration of the two-body bound states.
We can understand the origin of the negative differential conductance 
in the two-lead IRLM in terms of the formation of two-body bound states.
Such many-body bound states are also found in other open QD systems~\cite{%
Imamura-Nishino-Hatano_09PRB,Nishino-Imamura-Hatano_12JPCS}.

The present paper is organized as follows. 
In Sec.~\ref{sec:IRLM}, 
we introduce the two-lead IRLM with asymmetric interactions. 
In Sec.~\ref{sec:Landauer}, the extension of the Landauer formula~\cite{%
Nishino-Imamura-Hatano_09PRL,Nishino-Imamura-Hatano_11PRB} 
is described in a general setting.
In Sec.~\ref{sec:scattering-state}, 
we present the construction of the exact one- and two-electron 
scattering eigenstates
whose incident states are free-electronic plane waves in the leads.
We also give the $N$-electron scattering eigenstates
in the first order of the Coulomb-interaction parameters.
In Sec.~\ref{sec:electric-current}, 
through the extension of the Landauer formula,
we calculate the average electric current
for the system under finite bias voltages.
We obtain the universal electric current
by dealing with the divergences in the average electric current
with the Callan-Symanzik equation.
As a result, we observe the negative differential conductance
and the restoration of the suppressed electric current.
Section~\ref{sec:concluding-remarks} is devoted
to concluding remarks.

\section{Models and formulation}

\subsection{Interacting resonant-level models}
\label{sec:IRLM}

\begin{figure}[t]
\includegraphics[width=60mm,clip]{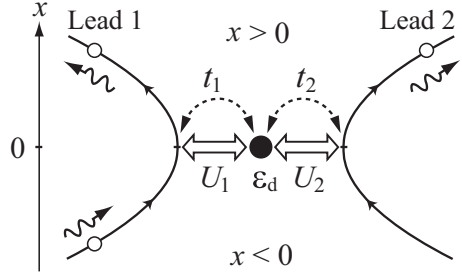}
\caption{\label{fig:IRLM-asym2-system} 
The two-lead interacting resonant-level model.
Electrons flow only upward due to the linearized dispersion
relations of the leads.
}
\end{figure}

We consider the open QD system described by the IRLM
of spinless electrons.
It consists of a QD with a single energy level
and two external leads of noninteracting electrons.
The arrangement of the QD and the two leads
is illustrated in Fig.~\ref{fig:IRLM-asym2-system}.
We assume the situation in which the QD is connected to
the two leads asymmetrically.

The Hamiltonian is given by
\begin{align}
\label{eq:IRLM}
H=
&\sum_{\ell=1,2}\int_{-\frac{L}{2}}^{\frac{L}{2}}\hspace{-3pt}
  dx\, c^{\dagger}_{\ell}(x)\frac{1}{\i}\frac{d}{dx}c_{\ell}(x)
 +\sum_{\ell=1,2}
  \big(t_{\ell}c^{\dagger}_{\ell}(0)d
  +t^{\ast}_{\ell}d^{\dagger}c_{\ell}(0)\big)
 +\epsilon_{{\rm d}}d^{\dagger}d
 \nn\\
&+\sum_{\ell=1,2}U_{\ell}
 c^{\dagger}_{\ell}(0)c_{\ell}(0)d^{\dagger}d.
\end{align}
Here $c^{\dagger}_{\ell}(x)$ and $c_{\ell}(x)$
are the creation- and the annihilation-operators of 
an electron in the lead $\ell$, while
$d^{\dagger}$ and $d$ are those on the QD.
The first term corresponds to the kinetic energy
of electrons in the leads, 
where $L$ stands for the length of the two leads
to be eventually taken to infinity.
The second term expresses the tunneling between the leads and the QD,
where the parameter $t_{\ell}(\in\mathbb{C})$ is the transfer integral.
We assume a single energy level $\epsilon_{\rm d}(\in\mathbb{R})$ on the QD,
which corresponds to the third term.
The fourth term describes the Coulomb interaction 
between the two electrons at the origin $x=0$ 
of the lead $\ell$ and on the QD.
The parameter $U_{\ell}(>0)$ is the strength of the Coulomb repulsion.

We focus on the electrons with positive velocities
in the vicinity of the Fermi energy $\epsilon_{\mathrm{F}}$ of each lead
and linearize the local dispersion relation to be 
$\epsilon(k)=v_{\mathrm{F}}(k-k_{\mathrm{F}})+\epsilon_{\mathrm{F}}$
under the assumption that the other parameters $|t_{\ell}|$, $|\epsilon_{\rm d}|$ 
and $U_{\ell}$ are small compared with the Fermi energy 
$\epsilon_{\mathrm{F}}$~\cite{Filyov-Wiegmann_80PLA}.
For simplicity, we have set $v_{\mathrm{F}}=1$,
$k_{\mathrm{F}}=0$ and $\epsilon_{\mathrm{F}}=0$ 
in Eq.~\eqref{eq:IRLM}.
Then, as is indicated in Fig.~\ref{fig:IRLM-asym2-system},
an electron coming from the part $x<0$ of the lead 1 
is scattered at the QD to the parts $x>0$ of the two leads.

In constructing the scattering eigenstates, we treat the system
as an open system in the limit $L\to\infty$ of the two leads.
In the sprit of the original Landauer formula~\cite{Datta,Imry}, 
we suppose that the infinite two leads can substitute for electron reservoirs
that are in the Fermi degenerate states of noninteracting electrons.
We assume that the electrons emitted from the electron reservoir into the lead $\ell$
follow the Fermi distribution function 
$f_{\mu_{\ell},\beta_{\ell}}(E)=1/(1+\e^{\beta_{\ell}(E-\mu_{\ell})})$
with the chemical potential $\mu_{\ell}$ 
and the inverse temperature $\beta_{\ell}$.
We are interested in the nonequilibrium steady state
realized between the large two electron reservoirs
in the cases $\mu_{1}\neq\mu_{2}$ of
different chemical potentials.

We adopt a standard definition of the electric-current operator as
\begin{align}
\label{eq:current_IRLM}
&I=\alpha I_{1}-(1-\alpha)I_{2},
 \quad
 I_{\ell}
 =\i\big(t_{\ell}c_{\ell}^{\dagger}(0)d
 -t^{\ast}_{\ell}d^{\dagger}c_{\ell}(0)\big).\quad
 (\ell=1,2)
\end{align}
For an arbitrary eigenstate $|\psi\rangle$
of the Hamiltonian $H$,
the expectation value $\langle\psi|I|\psi\rangle$
does not depend on the parameter $\alpha$
since the relation $\langle\psi|I_{1}|\psi\rangle
=-\langle\psi|I_{2}|\psi\rangle$ holds. 
In what follows, 
we choose the parameter $\alpha=|t_{2}|^{2}/t^{2}$
with $t=\sqrt{|t_{1}|^{2}+|t_{2}|^{2}}$
for the convenience of calculations.

\subsection{Extension of the Landauer formula}
\label{sec:Landauer}

Our purpose is to study the average electric current flowing  
across the QD of the two-lead IRLM beyond the linear-response regime.
The extension of the Landauer formula,
which was proposed in Refs.~\onlinecite{Nishino-Imamura-Hatano_09PRL}
and \onlinecite{Nishino-Imamura-Hatano_11PRB},
consists of the following three steps:
\begin{itemize}
\item[(i)] Construction of many-electron scattering eigenstates 
whose incident states are free-electronic plane waves in the leads;
\item[(ii)] Calculation of the quantum-mechanical expectation value of 
the electric current with the many-electron scattering eigenstates;
\item[(iii)] Calculation of the statistical-mechanical average of 
the electric current by assuming the equilibration of electrons
in each electron reservoir.
\end{itemize}
The many-electron scattering eigenstates constructed in the step (i) 
are characterized by the wave numbers of the incident plane waves.
We note that they are essentially different from the Bethe-ansatz 
eigenstates~\cite{Mehta-Andrei_06PRL,Filyov-Wiegmann_80PLA}, 
whose incident states are not free electronic
but include the effect of interactions.
The Bethe-ansatz result in Ref.~\onlinecite{Mehta-Andrei_06PRL} 
did not agree with results of the previous 
works~\cite{Doyon_07PRL,Golub_07PRB,Boulat-Saleur-Schmitteckert_08PRL}
while our results agree with them.
For the wave numbers $\{k_{1},\ldots,k_{N_{1}}\}$ 
of the $N_{1}$-electron incident plane wave 
coming in through the lead 1 and $\{h_{1},\ldots,h_{N_{2}}\}$ 
of the $N_{2}$-electron incident plane wave 
coming in through the lead 2,
we express the $N$-electron scattering eigenstates by $|k, h\rangle
=|k_{1},\ldots,k_{N_{1}},h_{1},\ldots,h_{N_{2}}\rangle$
with $N=N_{1}+N_{2}$.
In the step (ii), we calculate the expectation value 
$\langle k, h|I|k, h\rangle/\langle k, h|k, h\rangle$
of the electric-current operator $I$,
which we call the $N$-electron current.
This calculation is practically carried out by using 
the explicit $N$-electron scattering eigenstates.
In the step (iii), we take the limit $N_{\ell}, L\to\infty$  
of the $N$-electron current by assuming that  
the wave numbers $k_{i}$ and $h_{i}$ of incident plane waves
follow the Fermi distribution of each electron reservoir.
We call the limit an electron-reservoir limit.
Clearly, the reservoir limit corresponds to taking the 
statistical-mechanical average of the electric current 
for all the incident states that follow the Fermi distributions.
In general, the electrons scattered at the QD are 
in many-body states including the effect of interactions.
We assume that such many-body states are completely equilibrated
to the Fermi degenerate state of free electrons
in each election reservoir before being re-emitted towards the QD, 
which is the main assumption of the extension of the Landauer formula.
We shall see for the two-lead IRLM that, since the $N$-dependence of 
the $N$-electron current appears only in the upper bounds of the sums
on wave numbers $\{k_{i}\}$ and $\{h_{i}\}$, we can
take the reservoir limit by replacing the sums with the integrals 
on $k$ and $h$ with the Fermi-distribution functions 
$f_{\mu_{1},\beta}(k)$ and $f_{\mu_{2},\beta}(h)$.

The way of realizing the nonequilibrium steady states
in the extension of the Landauer formula is different from that in the Keldysh
formalism~\cite{%
Caroli-Combescot-Nozieres-SaintJames_71JPC,%
Meir-Wingreen-Lee_91PRL,Meir-Wingreen_92PRL,%
Hershfield-Davies-Wilkins_92PRB,Haug-Jauho,Golub_07PRB}.
In our extension of the Landauer formula, 
we first construct the $N$-electron scattering eigenstates 
for finite $N$ without the information 
of the equilibrium states in the electron reservoirs.
After the calculation of the $N$-electron current, we take 
the reservoir limit to consider the nonequilibrium steady state.
In the Keldysh formalism, on the other hand, the Green's functions
or the density operator describing the nonequilibrium steady states 
are obtained by adiabatically turning on the perturbative terms
for the initial nonperturbative steady states of
infinite number of electrons. 

Our approach is also independent of Hershfield's bias-operator approach, which 
constructs the density operator of the nonequilibrium steady states 
directly from one-electron field operators 
in the framework of the quantum field theory~\cite{%
Hershfield_93PRL,Han_06PRB,Anders_08PRL}.
We remark that the construction of the density operator 
through the bias-operator approach has not been established
analytically in interacting cases~\cite{Han_06PRB,Anders_08PRL}
except for the Toulouse limit of the Kondo model~\cite{%
Schiller-Hershfield_95PRB,Schiller-Hershfield_98PRB}. 

\section{Many-electron scattering eigenstates}
\label{sec:scattering-state}

\subsection{One-electron cases}

The linearization of the local dispersion relations of the leads
enables us to construct exact scattering eigenstates.
First, we consider the one-electron cases.
The one-electron scattering eigenstates are given in the form
\begin{align}
\label{eq:1-state_IRLM_1}
 |E\rangle
&=\Big(\int\!\!dx 
  \sum_{m=1,2}g_{m}(x)c^{\dagger}_{m}(x)
  +ed^{\dagger}\Big)|0\rangle,
\end{align}
where $|0\rangle$ is the vacuum state
satisfying $c_{\ell}(x)|0\rangle=d|0\rangle=0$.
The eigenfunctions $g_{m}(x)$ and $e$ are determined
by the coupled Schr\"odinger equations:
\begin{align}
\label{eq:1-state_IRLM_Sch-eq}
&\Big(\frac{1}{\i}\frac{d}{dx}-E\Big)g_{m}(x)
+t_{m}e\delta(x)=0, 
\quad (m=1,2) \nn\\
&(\epsilon_{{\rm d}}-E)e
+\sum_{m=1,2}t^{\ast}_{m}g_{m}(0)=0.
\end{align}
It is readily found that the eigenfunction $g_{m}(x)$
is discontinuous at $x=0$ and the matching condition 
at $x=0$ is obtained by
integrating the first equation in 
Eqs.~\eqref{eq:1-state_IRLM_Sch-eq} around $x=0$ as
\begin{align}
&g_{m}(0+)-g_{m}(0-)+\i t_{m}e=0.
\end{align}
Since the value $g_{m}(0)$ is not determined by the Schr\"odinger equations,
we assume $g_{m}(0)=(g_{m}(0+)+g_{m}(0-))/2$ from physical intuition.

To employ the Landauer formula, we need the scattering eigenstates
whose incident states are a plane wave in the lead 1 or the lead 2.
For the incident plane wave with the wave number $k$ in the lead $\ell$,
we consider the solution $g^{(\ell)}_{m,k}(x)$ and $e^{(\ell)}_{k}$
of the Schr\"odinger equations~\eqref{eq:1-state_IRLM_Sch-eq}
satisfying
\begin{align}
\label{eq:1-state_IRLM_sbc}
 g^{(\ell)}_{m,k}(x)
 =\frac{1}{\sqrt{2\pi}}
 \delta_{m\ell}\e^{\i kx}\quad
 \text{for}\; x<0,
\end{align}
where $\delta_{m\ell}$ is the Kronecker delta.
We refer to Eq.~\eqref{eq:1-state_IRLM_sbc} as scattering boundary conditions.
The solution with energy eigenvalue $E=k$ is given by
\begin{align}
\label{eq:1-state_IRLM_2}
&g^{(\ell)}_{m,k}(x)
 =\frac{1}{\sqrt{2\pi}}
 \big(\delta_{m\ell}-
 \i t_{m}\sqrt{2\pi}e^{(\ell)}_{k}\theta(x)\big)
 \e^{\i kx},
 \nn\\
&e^{(\ell)}_{k}
 =\frac{1}{\sqrt{2\pi}}\frac{t_{\ell}^{\ast}}
 {k-\epsilon_{{\rm d}}+\i\Gamma},
\end{align}
where $\theta(x)$ is the step function and 
$\Gamma=(|t_{1}|^{2}+|t_{2}|^{2})/2$ is the level width of the QD.
By inserting them into Eq.~\eqref{eq:1-state_IRLM_1}, 
we obtain the scattering eigenstate $|k;\ell\rangle$ whose 
incident state is a plane wave with wave number $k$ in the lead $\ell$.

The one-electron scattering eigenstates 
$|k;\ell\rangle$ 
are normalized on the $\delta$-function as 
$\langle k;\ell|k^{\prime};\ell^{\prime}\rangle
 =\delta_{\ell\ell^{\prime}}\delta(k-k^{\prime})$
in the limit $L\to\infty$.
In the calculation of quantum-mechanical expectation values
of physical quantities with the scattering eigenstates
$|k;\ell\rangle$, 
we need to restore the length $L$
of the leads in order to regularize the square norm as
$\langle k;\ell|k;\ell\rangle=L/(2\pi)$.

\subsection{Two-electron cases}

We next consider the two-electron cases
as the simplest example of the interacting cases.
The form of the two-electron scattering eigenstates
is given by
\begin{align}
|E\rangle
&=\Big(\sum_{l, m=1,2}
  \int_{x_{1}<x_{2}}\hspace{-10pt}dx_{1}dx_{2}\,
  g_{lm}(x_{1},x_{2})
  c^{\dagger}_{l}(x_{1})c^{\dagger}_{m}(x_{2})
 +\sum_{l=1,2}\int\!dx\,
  e_{l}(x)c^{\dagger}_{l}(x)d^{\dagger}
 \Big)
 |0\rangle.
\end{align}
Here we impose the antisymmetric relation 
$g_{lm}(x_{1},x_{2})=-g_{ml}(x_{2},x_{1})$.
The eigenvalue problem $H|E\rangle=E|E\rangle$ 
leads to the coupled Schr\"odinger equations:
\begin{subequations}
\begin{align}
\label{eq:2-elec_IRLM_Sch-eq_g}
&\Big(\frac{1}{\i}\Big(\frac{\partial}{\partial x_{1}}
 +\frac{\partial}{\partial x_{2}}\Big)-E\Big)
 g_{lm}(x_{1},x_{2})
 +t_{m}e_{l}(x_{1})\delta(x_{2})
 -t_{l}\delta(x_{1})e_{m}(x_{2})=0, 
 \\
\label{eq:2-elec_IRLM_Sch-eq_e}
&\Big(\frac{1}{\i}\frac{d}{dx}+\epsilon_{{\rm d}}
 +U_{l}\delta(x)-E\Big)e_{l}(x)
 +\hspace{-3pt}\sum_{m=1,2}
 t^{\ast}_{m}g_{lm}(x,0)=0.
\end{align}
\end{subequations}
In the previous works~\cite{%
Nishino-Imamura-Hatano_09PRL,Nishino-Imamura-Hatano_11PRB}
for the symmetric case $U_{1}=U_{2}$,
we employed the even-odd transformation that maps 
the two-lead IRLM to two single-lead systems.
However, since the transformation does not work 
for the asymmetric cases $U_{1}\neq U_{2}$,
we deal with the two-lead IRLM directly.

We present a construction of the exact two-electron 
scattering eigenstates, which is an extension 
of the previous one~\cite{Shen-Fan_07PRL,%
Nishino-Imamura-Hatano_09PRL,Nishino-Imamura-Hatano_11PRB}.
First, we derive three important 
relations from the Schr\"odinger equations
\eqref{eq:2-elec_IRLM_Sch-eq_g} 
and \eqref{eq:2-elec_IRLM_Sch-eq_e}.
The eigenfunction $g_{lm}(x_{1},x_{2})$
is discontinuous at $x_{1}=0$ and $x_{2}=0$,
while $e_{l}(x)$ is discontinuous at $x=0$.
The matching conditions at the discontinuous points
are given by
\begin{subequations}
\begin{align}
\label{eq:2-elec_IRLM_matching_g}
&g_{lm}(x,0+)-g_{lm}(x,0-)
 +\i t_{m}e_{l}(x)=0, 
 \\
\label{eq:2-elec_IRLM_matching_e}
&e_{l}(0+)-e_{l}(0-)+\i U_{l}e_{l}(0)=0,
\end{align}
\end{subequations}
which are obtained by integrating the Schr\"odinger equations
\eqref{eq:2-elec_IRLM_Sch-eq_g} and \eqref{eq:2-elec_IRLM_Sch-eq_e}
around the discontinuous points.
Since the values of the eigenfunctions at the discontinuous points
are not determined by the Schr\"odinger equations,
we assume 
\begin{subequations}
\begin{align}
\label{eq:2-elec_IRLM_disc_g}
&g_{lm}(x,0)
 =\frac{1}{2}\big(g_{lm}(x,0+)+g_{lm}(x,0-)\big),
 \\
\label{eq:2-elec_IRLM_disc_e}
&e_{l}(0)
 =\frac{1}{2}\big(e_{l}(0+)+e_{l}(0-)\big)
\end{align}
\end{subequations}
in a way similar to the one-electron cases.
By applying Eqs.~\eqref{eq:2-elec_IRLM_matching_g}
and \eqref{eq:2-elec_IRLM_disc_g} to 
Eq.~\eqref{eq:2-elec_IRLM_Sch-eq_e} for $x\neq 0$, we have
\begin{align}
&\Big(\frac{1}{\i}\frac{d}{dx}+\epsilon_{\rm d}-\i\Gamma-E\Big)
 e_{l}(x)
 =-\sum_{m}t_{m}^{\ast}g_{lm}(x,0-).
\end{align}
Given functions $g_{lm}(x,0-)$, $(m=1, 2)$, we obtain
the general solution for $e_{l}(x)$ as
\begin{align}
\label{eq:2-elec_IRLM_general-e}
e_{l}(x)
&=C_{l}\e^{\i(E-\epsilon_{\rm d}+\i\Gamma)x}
 -\i\sum_{m}t_{m}^{\ast}
 \int_{x_{0}}^{x}\hspace{-6pt}dz\,
 \e^{\i(E-\epsilon_{\rm d}+\i\Gamma)(x-z)}g_{lm}(z,0-),
\end{align}
where $C_{l}$ is the integration constant
and $x_{0}$ is chosen as $x_{0}=-\infty$ if $x<0$
and $x_{0}=0$ otherwise. 
On the other hand, by applying Eq.~\eqref{eq:2-elec_IRLM_disc_e} to 
Eq.~\eqref{eq:2-elec_IRLM_matching_e}, we have
the matching condition
\begin{align}
\label{eq:2-elec_IRLM_matching_e_2}
&\Big(1+\frac{\i}{2}U_{l}\Big)e_{l}(0+)
 =\Big(1-\frac{\i}{2}U_{l}\Big)e_{l}(0-).
\end{align}
The equations~\eqref{eq:2-elec_IRLM_matching_g},
\eqref{eq:2-elec_IRLM_general-e} and
\eqref{eq:2-elec_IRLM_matching_e_2}
are the relations that we need.

\def\arraystretch{1.3}
\begin{table}[t]
\begin{tabular}{|c|cc|cc|cc|}
\hline
 $(\ell_{1},\ell_{2})$
 & $A_{11,(12)}$ & $A_{11,(21)}$ 
 & $A_{12,(12)}$ & $A_{12,(21)}$ & $A_{22,(12)}$ & $A_{22,(21)}$ \\
\hline
 $(1,1)$
 & $1$ & $-1$ & $0$ & $0$ & $0$ & $0$ \\
 $(1,2)$ 
 & $0$ & $0$ & $1$ & $0$ & $0$ & $0$ \\
 $(2,2)$ 
 & $0$ & $0$ & $0$ & $0$ & $1$ & $-1$ \\
\hline
\end{tabular}
\caption{The coefficients $A_{\ell m,P}$ 
of the incident plane-wave states
for $(\ell_{1},\ell_{2})=(1,1), (1,2)$ and $(2,2)$.}
\label{tab:2-elec_incident}
\end{table}

Next, we demonstrate how to construct the two-electron
scattering eigenstates by the repeated use of 
the three equations \eqref{eq:2-elec_IRLM_matching_g},
\eqref{eq:2-elec_IRLM_general-e} and
\eqref{eq:2-elec_IRLM_matching_e_2}. 
Let us consider the situation
in which one electron with wave number $k_{1}$ 
is coming in through the lead $\ell_{1}$
and another with $k_{2}$ is coming in through the lead $\ell_{2}$.
We construct the eigenfunctions with energy eigenvalue
$E=k_{1}+k_{2}$ that satisfy
the scattering boundary conditions
\begin{align}
\label{eq:2-elec_IRLM_incident}
 g_{lm}(x_{1},x_{2})
&=\frac{1}{2\pi}\sum_{P}A_{lm,P}\e^{\i(k_{P_{1}}x_{1}+k_{P_{2}}x_{2})}
\end{align}
for $x_{1}, x_{2}<0$.
Here $P=(P_{1},P_{2})$ is a permutation of $(1,2)$ 
and the coefficients $A_{lm, P}$ are given by
$A_{lm,P}=\sgn(P)\delta_{l\ell_{P_{1}}}\delta_{m\ell_{P_{2}}}$
with the signature $\sgn(P)$ of the permutation $P$.
The coefficients $A_{lm,P}$ are explicitly listed
on Table~\ref{tab:2-elec_incident}.
Beginning from the incident state $g_{\ell m}(x_{1},x_{2})$ 
in the region $x_{1}<x_{2}<0$, we connect it to the other regions
through Eqs.~\eqref{eq:2-elec_IRLM_matching_g},
\eqref{eq:2-elec_IRLM_general-e} and
\eqref{eq:2-elec_IRLM_matching_e_2}.
By inserting $g_{lm}(x,0-)$ into 
Eq.~\eqref{eq:2-elec_IRLM_general-e}, we have
\begin{align}
\label{eq:2-elec_IRLM_e_1}
e_{l}(x)
&=\frac{1}{\sqrt{2\pi}}\sum_{P, m}
  A_{lm,P}e^{(m)}_{k_{P_{2}}}\e^{\i k_{P_{1}}x}
\end{align}
for $x<0$. Here we have set $x_{0}=-\infty$ and
have taken $C_{l}=0$ to avoid the divergence as $x\to -\infty$.
The function  $g_{lm}(x,0-)$ is connected to $g_{lm}(x,0+)$ 
by Eq.~\eqref{eq:2-elec_IRLM_matching_g} with 
Eq.~\eqref{eq:2-elec_IRLM_e_1} as
\begin{align}
&g_{lm}(x,0+)
 =\frac{1}{2\pi}\sum_{P, n}A_{ln,P}\e^{\i k_{P_{1}}x}
 (\delta_{mn}-\i t_{m}\sqrt{2\pi}e^{(n)}_{k_{P_{2}}}),
\end{align}
which leads to
\begin{align}
&g_{lm}(x_{1},x_{2})
 =\frac{1}{\sqrt{2\pi}}\sum_{P, n}A_{ln,P}
 \e^{\i k_{P_{1}}x_{1}}g_{m,k_{P_{2}}}^{(n)}(x_{2})
\end{align}
for $x_{1}<0<x_{2}$.
Recall that $g_{m,k}^{(n)}(x)$ is the one-electron
scattering eigenfunction in Eqs.~\eqref{eq:1-state_IRLM_2}.
Again, by inserting $g_{lm}(x,0-)$ into 
Eq.~\eqref{eq:2-elec_IRLM_general-e}, we have
\begin{align}
\label{eq:2-elec_IRLM_e_2}
e_{l}(x)
&=C^{\prime}_{l}\e^{\i(E-\epsilon_{\rm d}+\i\Gamma)x}
 -\sum_{P, m, n}A_{mn,P}
 g_{l,k_{P_{2}}}^{(n)}(x)
 e_{k_{P_{1}}}^{(m)}
\end{align}
for $x>0$. Here we keep the first term with 
the integration constant $C^{\prime}_{l}$
since the term is not divergent as $x\to\infty$.
By inserting $g_{lm}(0-,x)$ and $e_{l}(x)$, $(x>0)$
into Eq.~\eqref{eq:2-elec_IRLM_matching_g}
with the antisymmetric relation $g_{lm}(x_{1},x_{2})=-g_{ml}(x_{2},x_{1})$,
we have
\begin{align}
g_{lm}(x_{1},x_{2})
&=\sum_{P, r, n}A_{rn,P}
 g_{l, k_{P_{1}}}^{(r)}(x_{1})
 g_{m,k_{P_{2}}}^{(n)}(x_{2})
 +\i t_{l}C^{\prime}_{m}
 \e^{\i((\epsilon_{\rm d}-\i\Gamma)x_{1}+(E-\epsilon_{\rm d}+\i\Gamma)x_{2})}
\end{align}
for $0<x_{1}<x_{2}$.
Finally, by inserting Eqs.~\eqref{eq:2-elec_IRLM_e_1}
and \eqref{eq:2-elec_IRLM_e_2} into Eq.~\eqref{eq:2-elec_IRLM_matching_e_2},
we determine the integration constant $C_{l}^{\prime}$ as
\begin{align}
&C^{\prime}_{l}
 =\frac{1}{\sqrt{2\pi}}\i u_{l}
 \big(\delta_{l\ell_{2}}e^{(\ell_{1})}_{k_{1}}
 -\delta_{l\ell_{1}}e^{(\ell_{2})}_{k_{2}}\big)
\end{align}
with $u_{l}=2U_{l}/(2+\i U_{l})$.
Thus the two-electron scattering eigenfunctions satisfying
the scattering boundary conditions~\eqref{eq:2-elec_IRLM_incident}
are obtained as follows:
\begin{align}
\label{eq:2-elec_IRLM_eigen}
g^{(\ell_{1}\ell_{2})}_{lm,k_{1}k_{2}}(x_{1},x_{2})
&=g^{(\ell_{1})}_{l,k_{1}}(x_{1})
  g^{(\ell_{2})}_{m,k_{2}}(x_{2})
 -g^{(\ell_{2})}_{l,k_{2}}(x_{1})
  g^{(\ell_{1})}_{m,k_{1}}(x_{2})
 \nn\\
&\quad
 +t_{l}u_{m}
 Z^{(\ell_{1}\ell_{2})}_{m,k_{1}k_{2}}(x_{12})
 \e^{\i E x_{2}}\theta(x_{21})\theta(x_{1})
 -t_{m}u_{l}
 Z^{(\ell_{1}\ell_{2})}_{l,k_{1}k_{2}}(x_{21})
 \e^{\i E x_{1}}\theta(x_{12})\theta(x_{2}),
 \nn\\
e^{(\ell_{1}\ell_{2})}_{l,k_{1}k_{2}}(x)
&=g^{(\ell_{1})}_{l,k_{1}}(x)e^{(\ell_{2})}_{k_{2}}
 -g^{(\ell_{2})}_{l,k_{2}}(x)e^{(\ell_{1})}_{k_{1}}
 -\i u_{l}
 Z^{(\ell_{1}\ell_{2})}_{l,k_{1}k_{2}}(-x)
 \e^{\i Ex}\theta(x),
\end{align}
where $x_{ij}=x_{i}-x_{j}$ and
\begin{align}
\label{eq:Z-function}
 Z^{(\ell_{1}\ell_{2})}_{m,k_{1}k_{2}}(x)
 =\frac{1}{\sqrt{2\pi}}
 (\delta_{m\ell_{1}}e^{(\ell_{2})}_{k_{2}}
 -\delta_{m\ell_{2}}e^{(\ell_{1})}_{k_{1}})
 \e^{\i(\epsilon_{\rm d}-\i\Gamma)x}.
\end{align}
Here, on the left-hand sides of Eqs.~\eqref{eq:2-elec_IRLM_eigen} 
and \eqref{eq:Z-function},
we write the wave numbers $k_{1}$ and $k_{2}$
and the superscripts $\ell_{1}$ and $\ell_{2}$ of the leads
explicitly.

\begin{figure}[t]
\includegraphics[width=90mm,clip]{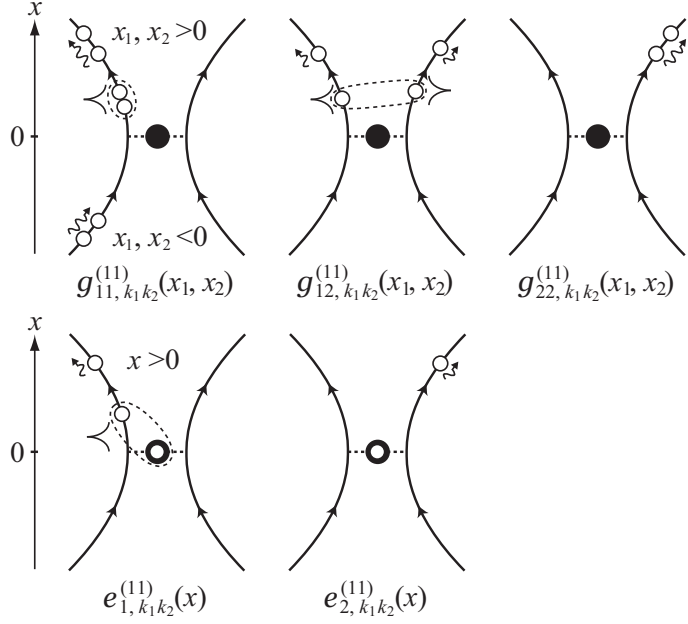}
\caption{\label{fig:IRLM-asym2-2BS} 
The two-electron scattering eigenfunctions
for the incident two electrons coming in through the lead 1.
The dotted circles surrounding the two electrons
indicate the two-body bound states,
which appear only when at least one of the electrons is reflected.
}
\end{figure}

Each term of the eigenfunctions in 
Eqs.~\eqref{eq:2-elec_IRLM_eigen} is interpreted as follows.
The first two terms correspond to the two-electron scattering 
eigenfunctions of the noninteracting cases,
which are given by the Slater determinant of the one-electron
scattering eigenfunctions in Eq.~\eqref{eq:1-state_IRLM_2}.
The effects of the interactions appear in the terms with 
the function $Z^{(\ell_{1}\ell_{2})}_{m,k_{1}k_{2}}(x)$.
They are interpreted as {\it two-body bound states} since 
they decay exponentially as the two electrons separate from each other.
For example, the third and the fourth terms in the eigenfunction 
$g^{(\ell_{1}\ell_{2})}_{lm,k_{1}k_{2}}(x_{1},x_{2})$
in Eq.~\eqref{eq:2-elec_IRLM_eigen} are rewritten as
\begin{align}
&t_{l}u_{m}
 Z^{(\ell_{1}\ell_{2})}_{m,k_{1}k_{2}}(x_{12})
 \e^{\i E x_{2}}\theta(x_{21})\theta(x_{1})
 -t_{m}u_{l}
 Z^{(\ell_{1}\ell_{2})}_{l,k_{1}k_{2}}(x_{21})
 \e^{\i E x_{1}}\theta(x_{12})\theta(x_{2})
 \nn\\
&=\big(t_{l}u_{m}
 Z^{(\ell_{1}\ell_{2})}_{m,k_{1}k_{2}}(0)
 \theta(x_{21})
 -t_{m}u_{l}
 Z^{(\ell_{1}\ell_{2})}_{l,k_{1}k_{2}}(0)
 \theta(x_{12})\big)
 \e^{(\i(\frac{E}{2}-\epsilon_{\rm d})-\Gamma)|x_{1}-x_{2}|%
 +\i E\frac{x_{1}+x_{2}}{2}}
 \theta(x_{1})\theta(x_{2}).
\end{align}
The binding length of the two-body bound states is given by 
$1/\Gamma$ where $\Gamma$ is the level width of the QD.
It should be emphasized that the two-body bound states 
are characteristic to open systems and do not appear
under periodic boundary conditions.
We also find that the two-body bound states are associated with
the electron that is reflected at the QD.
For example, if the incident two electrons come in through the lead 1
($\ell_{1}=\ell_{2}=1$), 
the two-body bound states appear only in the eigenfunctions
$g^{(11)}_{11,k_{1}k_{2}}(x_{1},x_{2})$,
$g^{(11)}_{12,k_{1}k_{2}}(x_{1},x_{2})$
and $e^{(11)}_{1,k_{1}k_{2}}(x)$
since $Z^{(11)}_{2,k_{1}k_{2}}(x)=0$ 
as is depicted in Fig.~\ref{fig:IRLM-asym2-2BS}.

It is instructive to inspect the set of wave numbers
characterizing each term of the eigenfunctions in 
Eqs.~\eqref{eq:2-elec_IRLM_eigen}.
As is found from the terms of the two-body bound states,
the wave-number set $\{k_{1}, k_{2}\}$ of the incident states
is not conserved and is scattered to 
the set $\{\epsilon_{\rm d}-\i\Gamma,E-\epsilon_{\rm d}+\i\Gamma\}$
including the imaginary part $\i\Gamma$. 
We note that the terms of the two-body bound states 
decay with the distance of the two electrons 
but are stationary in time since the total energy eigenvalue is real;
the imaginary parts of the complex wave numbers cancel out each other 
in the total energy eigenvalue.
By the completeness of the plane-wave functions 
$\{\e^{\i(k_{1}x_{1}+k_{2}x_{2})}| k_{1}, k_{2}\in\mathbb{R}\}$,
the terms are expanded as
\begin{align}
&\e^{(\i(\frac{E}{2}-\epsilon_{\rm d})-\Gamma)|x_{1}-x_{2}|%
 +\i E\frac{x_{1}+x_{2}}{2}}
 =\int dk^{\prime}_{1}dk^{\prime}_{2}\frac{1}{E-E^{\prime}+\i 0}
 c(k^{\prime}_{1},k^{\prime}_{2};x_{1},x_{2})
 \e^{\i(k^{\prime}_{1}x_{1}+k^{\prime}_{2}x_{2})},
\end{align}
where $E^{\prime}=k^{\prime}_{1}+k^{\prime}_{2}$ and
$c(k^{\prime}_{1},k^{\prime}_{2};x_{1},x_{2})$ is the coefficient of the expansion.
Hence we find that the terms of the two-body bound states describe 
the various scattering processes to the sets $\{k^{\prime}_{1},k^{\prime}_{2}\}$
satisfying energy conservation $E=E^{\prime}$.
The coefficient $c(k^{\prime}_{1},k^{\prime}_{2};x_{1},x_{2})$
has the poles on the complex $k^{\prime}_{1}$- and $k^{\prime}_{2}$-planes 
which come from the resonant pole $k=\epsilon_{\rm d}-\i\Gamma$ 
of the one-electron scattering eigenfunctions in 
Eqs.~\eqref{eq:2-elec_IRLM_general-e}.
Thus the two-body bound states appear as a consequence of
one-body resonances~\cite{Ordonez-Nishino-Hatano_14preprint}.

The appearance of such many-body bound states is expected 
for general open QD systems with localized interactions.
We have shown that two-body bound states appear in the Anderson model 
with spin degrees of freedom~\cite{Imamura-Nishino-Hatano_09PRB}
and the double QD systems~\cite{Nishino-Imamura-Hatano_12JPCS}.
In Ref.~\onlinecite{Lebedev-Lesovik-Blatter_08PRL},
the $N$-electron scattering matrix for another QD system
with interactions was explicitly constructed in a real-time representation,
where two-electron scattering eigenstates are obtained in an integral form.
We speculate that, by evaluating the integral form, 
two-body bound states similar to ours should appear.

\mathversion{bold}
\subsection{$N$-electron cases}
\mathversion{normal}

We can obtain the exact $N$-electron scattering eigenstates for arbitrary $N$.
The scattering eigenstates for a few electrons can be constructed
in a way similar to the two-electron cases.
In the three-electron scattering eigenstates, for example,
two electrons out of the three form the two-body bound states 
after the scattering at the QD~\cite{Nishino-Imamura-Hatano_09PRL}.
The explicit form of the scattering eigenstates for a few electrons 
leads to a conjectural form of the $N$-electron scattering eigenstates.
We have shown that they are indeed the eigenstates.

We present only the results in the first order of $U_{\ell}$, 
which we need in the next section.
The form of the $N$-electron scattering eigenstates is given by
\begin{align}
\label{eq:N-elec_IRLM}
|E\rangle
&=\Big(\sum_{\{m_{i}\}}
  \int_{x_{1}<\cdots<x_{N}}\hspace{-20pt}d^{N}x\,
  g_{m_{1}\cdots m_{N}}(x_{1},\ldots,x_{N})
  c^{\dagger}_{m_{1}}(x_{1})\cdots c^{\dagger}_{m_{N}}(x_{N})
  \nn\\
&\quad
  +\sum_{\{n_{i}\}}
  \int_{x_{1}<\cdots<x_{N-1}}\hspace{-30pt}d^{N-1}x\,
  e_{n_{1}\cdots n_{N-1}}(x_{1},\ldots,x_{N-1})
  c^{\dagger}_{n_{1}}(x_{1})\cdots c^{\dagger}_{n_{N-1}}(x_{N-1})
  d^{\dagger}
  \Big)
 |0\rangle.
\end{align}
Here we impose the antisymmetric relations for 
the $N$-electron eigenfunctions as follows:
\begin{align}
&g_{m_{Q_{1}}\cdots m_{Q_{N}}}(x_{Q_{1}},\ldots,x_{Q_{N}})
 =\sgn(Q)g_{m_{1}\cdots m_{N}}(x_{1},\ldots,x_{N}),
 \nn\\
&e_{n_{R_{1}}\cdots n_{R_{N-1}}}
 (x_{R_{1}},\ldots,x_{R_{N-1}})
 =\sgn(R)e_{n_{1}\cdots n_{N-1}}(x_{1},\ldots,x_{N-1}),
\end{align}
where $Q=(Q_{1},\ldots,Q_{N})$ is a permutation of $(1,2,\ldots,N)$
and $R=(R_{1},\ldots,R_{N-1})$ 
is that of $(1,2,\ldots,N-1)$.
We consider the situation in which the electron 
with wave number $k_{i}, (i=1, \ldots, N)$
comes in through the lead $\ell_{i}$ to the QD.
The $N$-electron scattering eigenfunctions are constructed
in the first order of $U_{\ell}$ as
\begin{align}
\label{eq:N-elec_IRLM_eigen_g}
&g^{(\ell_{1}\cdots\ell_{N})}_{m_{1}\cdots m_{N},k_{1}\cdots k_{N}}
(x_{1},\ldots,x_{N})
 \nn\\
&=\sum_{P}\sgn(P)
 \prod_{i=1}^{N}
 g^{(\ell_{P_{i}})}_{m_{i},k_{P_{i}}}(x_{i})
 \nn\\
&\quad
 +\frac{1}{2}\sum_{P, Q}\sgn(PQ)
 \prod_{i=1}^{N-2}
 g^{(\ell_{P_{i}})}_{m_{Q_{i}},k_{P_{i}}}(x_{Q_{i}})
 \theta(x_{Q_{N-2}},\ldots,x_{Q_{1}})
 \nn\\
&\qquad\times
 t_{m_{Q_{N-1}}}U_{m_{Q_{N}}}
 Z^{(\ell_{P_{N-1}}\ell_{P_{N}})}_{%
 m_{Q_{N}},k_{P_{N-1}}k_{P_{N}}}(x_{Q_{N-1}Q_{N}})
 \e^{\i(k_{P_{N-1}}+k_{P_{N}})x_{Q_{N}}}\theta(x_{Q_{N}Q_{N-1}})
 \theta(x_{Q_{N-1}})
 \nn\\
&\quad
 +O(U^{2}),
\end{align}
where $P$ and $Q$ are permutations of $(1,2,\ldots,N)$, and
\begin{align}
\label{eq:N-elec_IRLM_eigen_e}
&e^{(\ell_{1}\cdots\ell_{N})}_{n_{1}\cdots n_{N-1},k_{1}\cdots k_{N}}
(x_{1},\ldots,x_{N-1})
 \nn\\
&=\sum_{P}\sgn(P)
 \prod_{i=1}^{N-1}
 g^{(\ell_{P_{i}})}_{n_{i},k_{P_{i}}}(x_{i})
 e^{(\ell_{P_{N}})}_{k_{P_{N}}}
 \nn\\
&\quad
 +\frac{1}{2}\sum_{P, Q}\sgn(PQ)
 \prod_{i=1}^{N-3}
 g^{(\ell_{P_{i}})}_{n_{Q_{i}},k_{P_{i}}}(x_{Q_{i}})
 e^{(\ell_{P_{N-2}})}_{k_{P_{N-2}}}
 \theta(x_{Q_{N-3}},\ldots,x_{Q_{1}})
 \nn\\
&\qquad\times
 t_{n_{Q_{N-2}}}U_{n_{Q_{N-1}}}
 Z^{(\ell_{P_{N-1}}\ell_{P_{N}})}_{%
 n_{Q_{N-1}},k_{P_{N-1}}k_{P_{N}}}(x_{Q_{N-2}Q_{N-1}})
 \e^{\i(k_{P_{N-1}}+k_{P_{N}})x_{Q_{N-1}}}
 \theta(x_{Q_{N-1}Q_{N-2}})
 \theta(x_{Q_{N-2}})
 \nn\\
&\quad
 -\frac{\i}{2}\sum_{P, Q}\sgn(PQ)
 \prod_{i=1}^{N-2}
 g^{(\ell_{P_{i}})}_{n_{Q_{i}},k_{P_{i}}}(x_{Q_{i}})
 \theta(x_{Q_{N-2}},\ldots,x_{Q_{1}})
 \nn\\
&\qquad\times
 U_{n_{Q_{N-1}}}
 Z^{(\ell_{P_{N-1}}\ell_{P_{N}})}_{%
 n_{Q_{N-1}},k_{P_{N-1}}k_{P_{N}}}(-x_{Q_{N-1}})
 \e^{\i(k_{P_{N-1}}+k_{P_{N}})x_{Q_{N-1}}}
 \theta(x_{Q_{N-1}})
 \nn\\
&\quad
 +O(U^{2}),
\end{align}
where $P$ is a permutation of $(1,2,\ldots,N)$
and $Q$ is that of $(1,2,\ldots,N-1)$.
Here we have used the notation
\begin{align}
&\theta(x_{m},\ldots,x_{2},x_{1})
 =\theta(x_{m,m-1})\cdots\theta(x_{32})\theta(x_{21}).
\end{align}
The third term in Eq.~\eqref{eq:N-elec_IRLM_eigen_e} 
corresponds to the configuration in which 
one of the two electrons that form the two-body bound states is on the QD.
On the other hand, the second term in Eq.~\eqref{eq:N-elec_IRLM_eigen_e} 
corresponds to the configuration in which both of
the two electrons that form the two-body bound states are in the leads,
which has not appeared in the two-electron scattering 
eigenfunction $e^{(\ell_{1}\ell_{2})}_{l,k_{1}k_{2}}(x)$ 
in Eqs.~\eqref{eq:2-elec_IRLM_eigen}
only in $g^{(\ell_{1}\ell_{2})}_{l,k_{1}k_{2}}(x_{1},x_{2})$ 
in Eqs.~\eqref{eq:2-elec_IRLM_eigen}.
In what follows, we denote 
the eigenstate obtained by inserting the eigenfunctions in 
Eqs.~\eqref{eq:N-elec_IRLM_eigen_g}
and \eqref{eq:N-elec_IRLM_eigen_e} into Eq.~\eqref{eq:N-elec_IRLM}
by $|k;\ell\rangle
=|k_{1},\ldots,k_{N}; \ell_{1},\ldots,\ell_{N}\rangle$.

\section{Electric current under bias voltages}
\label{sec:electric-current}

\mathversion{bold}
\subsection{$N$-electron current}
\mathversion{normal}

By following the three steps of the extension of the Landauer formula
given in Sec.~\ref{sec:Landauer},
we next calculate the average electric current
for the system under finite bias voltages~\cite{%
Nishino-Imamura-Hatano_09PRL,Nishino-Imamura-Hatano_11PRB}.
First, we calculate the $N$-electron current, that is,
the quantum-mechanical expectation value of the electric-current 
operator $I$ with the $N$-electron scattering eigenstates $|k;\ell\rangle$.
We assume $k_{i}\neq k_{j}$ if $\ell_{i}=\ell_{j}$ and
restrict our calculation to the first order of $U_{\ell}$.
We need to calculate the following overlap integral:
\begin{align}
\label{eq:N-elec_current_0}
\langle k;\ell|I|k;\ell\rangle
&=2\mathrm{Im}\Big(
  \int_{x_{1}<\cdots<x_{N-1}}\hspace{-30pt}
  d^{N-1}x
  \sum_{\ell, \{n_{i}\}}(-1)^{\ell}
  \frac{|t_{\Bar{\ell}}|^{2}}{t^{2}}t_{\ell}
  \nn\\
&\quad \times
  g^{\ast}_{n_{1}\ldots n_{N-1}\ell}(x_{1},\ldots,x_{N-1},0)
  e_{n_{1}\ldots n_{N-1}}(x_{1},\ldots,x_{N-1})
  \Big),
\end{align}
where $\Bar{\ell}=3-\ell$.
By inserting the $N$-electron eigenfunctions 
in Eqs.~\eqref{eq:N-elec_IRLM_eigen_g} 
and \eqref{eq:N-elec_IRLM_eigen_e} into 
Eq.~\eqref{eq:N-elec_current_0},
we obtain 
\begin{align}
\label{eq:N-elec_current-QM}
&\langle k;\ell|I|k;\ell\rangle
 \nn\\
&=\frac{|t_{1}t_{2}|^{2}}{\pi t^{2}}
 \Big(\frac{L}{2\pi}\Big)^{N-1}
 \sum_{i}(-1)^{\ell_{i}}
 \mathrm{Im}(G_{k_{i}})
 \nn\\
&\quad -\Big(\frac{L}{2\pi}\Big)^{N-2}
 \sum_{i\neq j}(-1)^{\ell_{i}}
 \sum_{m}U_{m}
 \mathrm{Im}\Big(\frac{2t_{1}t_{2}}{t^{2}}
 e^{(\Bar{\ell}_{i})}_{k_{i}}
 g^{(\ell_{j})\ast}_{m,k_{j}}(0+)
 Z^{(\ell_{i}\ell_{j})}_{\ell,k_{i}k_{j}}(0)
 \Big)
 \nn\\
&\quad 
 -\frac{2|t_{1}t_{2}|^{2}}{\pi t^{2}}
 \Big(\frac{L}{2\pi}\Big)^{N-3}
 \hspace{-5pt}\sum_{i\neq j\neq l(\neq i)}
 \hspace{-5pt}(-1)^{\ell_{i}}
 \mathrm{Im}(G_{k_{i}})
 \sum_{m}U_{m}
 \mathrm{Re}
 \big(
 G_{k_{j}}
 g^{(\ell_{l})\ast}_{m,k_{l}}(0+)
 e^{(\ell_{j})\ast}_{k_{j}}
 Z^{(\ell_{j}\ell_{l})}_{m,k_{j}k_{l}}(0)
 \big)
 \nn\\
&\quad 
 +O(U^{2}).
\end{align}
Here $L$ is the system length coming from
the regularized square norm $\langle k; \ell|k; \ell\rangle=L/(2\pi)$ 
of the one-electron scattering eigenstates. 
In order to express the results, 
we have used the notation 
\begin{align}
&G_{k}
 =\frac{1}{k-\epsilon_{{\rm d}}+\i\Gamma},
\end{align}
which is the one-electron Green's function on the QD.
We notice that the choice of the parameter $\alpha$ 
in Eqs.~\eqref{eq:current_IRLM} simplifies the calculation.
On the other hand, the square norm of the $N$-electron eigenstates
is calculated as
\begin{align}
\label{eq:N-elec_norm}
&\langle k;\ell|k;\ell\rangle
 \nn\\
&=\sum_{\{m_{i}\}}
 \int_{x_{1}<\cdots<x_{N}}\hspace{-20pt}
 d^{N}x\,|g_{m_{1}\cdots m_{N}}(x_{1},\ldots,x_{N})|^{2}
 +\sum_{\{n_{i}\}}
 \int_{x_{1}<\cdots<x_{N-1}}\hspace{-30pt}
 d^{N-1}x\,|e_{n_{1}\cdots n_{N-1}}(x_{1},\ldots,x_{N-1})|^{2}
 \nn\\
&=\Big(\frac{L}{2\pi}\Big)^{N}
 -2
 \Big(\frac{L}{2\pi}\Big)^{N-2}
 \sum_{i\neq j}\sum_{m=1,2}U_{m}
 \mathrm{Re}\big(G_{k_{i}}
 g^{(\ell_{j})\ast}_{m,k_{j}}(0+)
 e^{(\ell_{i})\ast}_{k_{i}}
 Z^{(\ell_{i}\ell_{j})}_{m,k_{i}k_{j}}(0)
 \big)
 +O(U^{2}).
\end{align}
It should be noted that the term in the $(N-1)$th order in $L$
does not appear above.
Combining Eqs.~\eqref{eq:N-elec_current-QM}
and \eqref{eq:N-elec_norm}, 
we obtain the $N$-electron current as
\begin{align}
\label{eq:current-QM_1}
&\frac{\langle k;\ell|I|k;\ell\rangle}
 {\langle k;\ell|k;\ell\rangle}
 \nn\\
&=\frac{|t_{1}t_{2}|^{2}}{\pi t^{2}}\frac{2\pi}{L}\sum_{i=1}^{N}
 (-1)^{\ell_{i}}\mathrm{Im}(G_{k_{i}})
 \nn\\
&\quad -\frac{4\pi^{2}}{L^{2}}
 \sum_{i\neq j}(-1)^{\ell_{i}}
 \sum_{m=1,2}U_{m}\mathrm{Im}\Big(\frac{2t_{1}t_{2}}{t^{2}}
 e^{(\Bar{\ell}_{i})}_{k_{i}}
 g^{(\ell_{j})\ast}_{m,k_{j}}(0+)
 Z^{(\ell_{i}\ell_{j})}_{m,k_{i}k_{j}}(0)
 \Big)
 \nn\\
&\quad +\frac{2|t_{1}t_{2}|^{2}}{\pi t^{2}}
 \frac{8\pi^{3}}{L^{3}}\sum_{i\neq j}
 \mathrm{Im}\big((-1)^{\ell_{i}}G_{k_{i}}
 +(-1)^{\ell_{j}}G_{k_{j}}\big)
 \sum_{m=1,2}U_{m}
 \mathrm{Re}\big(
 G_{k_{i}}
 g^{(\ell_{j})\ast}_{m,k_{j}}(0+)
 e^{(\ell_{i})\ast}_{k_{i}}
 Z^{(\ell_{i}\ell_{j})}_{m,k_{i}k_{j}}(0)
 \big)
 \nn\\
&\quad
 +O(U^{2}).
\end{align}

\subsection{Average electric current}
\label{sec:average-electric-current}

Next, we take the reservoir limit of 
the $N$-electron current in Eq.~\eqref{eq:current-QM_1}
to obtain the average electric current.
We assume that the infinite lead substitutes for
a large electron reservoir characterized 
by the Fermi distribution function 
$f_{\mu,\beta}(k)=1/(1+\e^{\beta(k-\mu)})$
with a chemical potential $\mu$
and an inverse temperature $\beta$.
We also assume that electrons are completely equilibrated 
in each electron reservoir 
before being re-emitted towards the QD,
which is the main assumption of the extension of the Landauer formula.

In the $N$-electron current in Eq.~\eqref{eq:current-QM_1},
the $N$-dependence appears only in the upper bounds 
of the sums on the wave numbers together with the factor $2\pi/L$,
which means that we can take the reservoir limit 
$N_{\ell}, L\to\infty$ described in Sec.~\ref{sec:Landauer}.
It should be noted that
the first term in Eq.~\eqref{eq:current-QM_1} 
contains a single sum on $i$ with the factor $2\pi/L$ and
the second term contains a double sum on $i$ and $j$
with $(2\pi/L)^{2}$ while
the third term is a double sum on $i$ and $j$
with $(2\pi/L)^{3}$ due to the square norm 
$\langle k;\ell|k;\ell\rangle$ appearing in the denominator. 
Therefore the third term in Eq.~\eqref{eq:current-QM_1}
vanishes in the reservoir limit $N_{\ell}, L\to\infty$.

In order to investigate the average electric current,
we set $\ell_{1}=\cdots=\ell_{N_{1}}=1$
and $\ell_{N_{1}+1}=\cdots=\ell_{N}=2$ 
in Eq.~\eqref{eq:current-QM_1} and relabel
$|k;\ell\rangle$ by $|k, h\rangle$ with
$h_{i}=k_{N_{1}+i}$, ($1\leq i\leq N_{2}$).
The $N$-electron current is rewritten as follows:
\begin{subequations}
\begin{align}
\label{eq:N-elec-current}
&\frac{\langle k, h|I|k, h\rangle}
 {\langle k, h|k, h\rangle}
 =J_{0}+J_{1}+J_{2}+O(U^{2}),
 \\
\label{eq:N-elec-current_J0}
&J_{0}=
 -\frac{2\Gamma_{1}\Gamma_{2}}{\pi \Gamma}
 \frac{2\pi}{L}\Big[\sum_{i=1}^{N_{1}}
 \mathrm{Im}\big(G_{k_{i}}\big)
 -\sum_{i=1}^{N_{2}}
 \mathrm{Im}\big(G_{h_{i}}\big)\Big],
 \\
\label{eq:N-elec-current_J1}
&J_{1}
 =-\frac{\Gamma_{1}\Gamma_{2}}{\pi^{2}\Gamma}
 \frac{4\pi^{2}}{L^{2}}\Big[
 U_{1}\hspace{-8pt}
 \sum_{1\leq i\neq j\leq N_{1}}\hspace{-8pt}
 \xi^{(11)}_{k_{i}k_{j}}
 +\hspace{-5pt}\sum_{1\leq i\leq N_{1}\atop 1\leq j\leq N_{2}}\hspace{-5pt}
 \big(U_{1}\xi^{(12)}_{k_{i}h_{j}}+U_{2}\xi^{(21)}_{h_{j}k_{i}}\big)
 +U_{2}\hspace{-8pt}\sum_{1\leq i\neq j\leq N_{2}}\hspace{-8pt}
 \xi^{(22)}_{h_{i}h_{j}}\Big],
\end{align}
\end{subequations}
where we use
\begin{align}
 \xi^{(lm)}_{k h}
&=\mathrm{Im}\Big[G_{h}
 \Big((-1)^{l}\delta_{lm}G_{k}-(-1)^{m}G_{h}
 +2\i\big((-1)^{l}\Gamma_{m}G^{\ast}_{h}G_{k}
 -(-1)^{m}\Gamma_{l}G^{\ast}_{k}G_{h}
 \big)
 \Big)\Big]
\end{align}
and $\Gamma_{l}=|t_{l}|^{2}/2$.
We omit the explicit form of $J_{2}$ in Eq.~\eqref{eq:N-elec-current}
since it does not contribute to the average electric current
in the reservoir limit $N_{\ell}, L\to\infty$.
Thus the parts of the $N$-electron current 
that contribute to the average electric current are detemined by 
the two-electron scattering eigenstates.

In the reservoir limit $N_{\ell}, L\to\infty$,
we replace the sums on $k_{i}$ in Eq.~\eqref{eq:N-elec-current}
by the integral on $k$ with $f_{\mu,\beta}(k)$ as
\begin{align}
&\frac{2\pi}{L}\sum_{i=1}^{N_{\ell}}J(k_{i})
 \to\int_{-\Lambda}^{\infty}\hspace{-5pt} 
 dk\, f_{\mu_{\ell},\beta_{\ell}}(k)J(k),
\end{align}
where we need to introduce the low-energy cutoff $-\Lambda$
since the local dispersion relation of the lead is bottomless. 
At zero temperature $(\beta_{\ell}=\infty)$, 
the average electric current is given by
\begin{align}
\label{eq:average-current_1}
\langle I\rangle
&=-\frac{2\Gamma_{1}\Gamma_{2}}{\pi\Gamma}
  \int_{\mu_{2}}^{\mu_{1}}\hspace{-8pt}dk\, 
  \mathrm{Im}\big(G_{k}\big)
  -\frac{\Gamma_{1}\Gamma_{2}}{\pi\Gamma}\sum_{\ell=1,2}U_{\ell}
   \int_{-\Lambda}^{\mu_{\ell}}\hspace{-10pt}dk\,
   \Big(\int_{-\Lambda}^{\mu_{\ell}}\hspace{-10pt}dh\, 
   \xi^{(\ell\ell)}_{kh}
  +\int_{-\Lambda}^{\mu_{\Bar{\ell}}}\hspace{-10pt}dh\, 
   \xi^{(\ell\Bar{\ell})}_{kh}\Big)
 +O(U^{2}).
\end{align}
We notice that the first term in Eq.~\eqref{eq:average-current_1}
reproduces the original Landauer formula in the noninteracting cases.
The double summations in Eq.~\eqref{eq:N-elec-current_J1}
give double integrals in the second term in Eq.~\eqref{eq:average-current_1},
which give a contribution of the Coulomb interactions.
Through the integral formulas
\begin{subequations}
\begin{align}
 \int_{-\Lambda}^{\mu_{\ell}}\hspace{-5pt}
 dk\,G_{k}
&=\frac{1}{2}\log
 \Big(\frac{\epsilon_{\ell}^{2}+1}{\epsilon_{\Lambda}^{2}+1}\Big)
 +\i(\arctan(\epsilon_{\ell})-\arctan(\epsilon_{\Lambda})),
 \nn\\
 \int_{-\Lambda}^{\mu_{\ell}}\hspace{-5pt}
 dk\,G_{k}^{2}
&=\frac{1}{\Gamma}\Big(\frac{1}{\epsilon_{\ell}-\i}
 -\frac{1}{\epsilon_{\Lambda}-\i}\Big),
 \\
 \int_{-\Lambda}^{\mu_{\ell}}\hspace{-5pt}
 dk\,G^{\ast}_{k}G_{k}
&=-\frac{1}{\Gamma}
 (\arctan(\epsilon_{\ell})-\arctan(\epsilon_{\Lambda})),
\end{align}
\end{subequations}
with $\epsilon_{\ell}=(\epsilon_{\rm d}-\mu_{\ell})/\Gamma$
and $\epsilon_{\Lambda}=(\epsilon_{\rm d}+\Lambda)/\Gamma$,
we obtain the average electric current
\begin{align}
\label{eq:average-current_2}
\langle I\rangle
&=-\frac{2\Gamma_{1}\Gamma_{2}}{\pi\Gamma}j_{-}
 \nn\\
&\quad
 +\frac{\Gamma_{1}\Gamma_{2}}{\pi^{2}\Gamma^{2}}
 \sum_{\ell=1,2}U_{\ell}
 \Big[\big(\Gamma_{\Bar{\ell}}j_{-}-\Gamma_{\ell}j_{1}\big)\log
 \Big(\frac{\epsilon_{\ell}^{2}+1}{\epsilon_{\Lambda}^{2}+1}\Big)
 +\big(\Gamma(\epsilon_{\ell}-\epsilon_{\Lambda})
 -\Gamma_{\ell}
 \big(2\arctan(\epsilon_{\ell})-j_{\Lambda}\big)\big)
 j_{2}\Big]
 \nn\\
&\quad
 +O(U^{2}),
\end{align}
where we use the notation
\begin{align}
&j_{-}=\arctan(\epsilon_{1})-\arctan(\epsilon_{2}),\quad
 j_{\Lambda}=2\arctan(\epsilon_{\Lambda}),
 \nn\\
&j_{s}=\frac{\epsilon_{1}^{2-s}}{\epsilon_{1}^{2}+1}
 -\frac{\epsilon_{2}^{2-s}}{\epsilon_{2}^{2}+1},\quad
 (s=1,2).
\end{align}
We find that the average electric current $\langle I\rangle$
contains linear and logarithmic divergences
in the limit $\Lambda\to\infty$, which is similar to
the symmetric case $U_{1}=U_{2}$~\cite{Doyon_07PRL,%
Nishino-Imamura-Hatano_09PRL,Nishino-Imamura-Hatano_11PRB}.

\subsection{Universal electric current}

We employ a renormalization-group technique to 
deal with the divergences in the average electric current
in Eq.~\eqref{eq:average-current_2}.
The divergences are due to the bottomless dispersion relation.
By the renormalization-group analysis, we zoom into the Fermi
energy and thereby discard all details that arise from
the specifics of the dispersion relation.
As a result, we obtain a universal form of 
the average electric current.

We devise a Callan-Symanzik equation~\cite{Doyon_07PRL,%
Doyon-Andrei_06PRB} 
so that the average electric current may satisfy it.
Let us introduce a parameter $D=\Gamma\sqrt{\epsilon_{\Lambda}^{2}+1}$.
We can indeed see that, for $D\gg\Gamma, |\epsilon_{\rm d}|$,
the average electric current $\langle I\rangle$
in Eq.~\eqref{eq:average-current_2} satisfies 
\begin{align}
\label{eq:Callan-Symanzik-eq}
&\Big(D\frac{\partial}{\partial D}
 +\sum_{\ell=1,2}
  \beta_{\Gamma_{\ell}}\frac{\partial}{\partial\Gamma_{\ell}}
 +\beta_{\epsilon_{\rm d}}\frac{\partial}{\partial\epsilon_{\rm d}}
 \Big)
 \langle I\rangle=0,
\end{align}
where the beta functions $\beta_{\Gamma_{\ell}}$ 
and $\beta_{\epsilon_{\rm d}}$ are given
in the first order of $U_{\ell}$ as
\begin{align}
&\beta_{\Gamma_{\ell}}
 =-\frac{U_{\ell}}{\pi}\Gamma_{\ell}+O(U^{2}),
 \quad
 \beta_{\epsilon_{\rm d}}
 =-\frac{\Bar{U}}{\pi}D+O(U^{2})
\end{align}
with the average interaction $\Bar{U}=(U_{1}+U_{2})/2$.
The Callan-Symanzik equation of the form~\eqref{eq:Callan-Symanzik-eq} 
is an extension of the previous ones ~\cite{Doyon_07PRL,%
Nishino-Imamura-Hatano_09PRL,Nishino-Imamura-Hatano_11PRB} 
for the case of the symmetric couplings.

The general solution of the Callan-Symanzik equation
determines a scaling form of the average electric current as
\begin{align}
 \langle I\rangle
 =J\Big(D^{\frac{U_{1}}{\pi}}\Gamma_{1},
 D^{\frac{U_{2}}{\pi}}\Gamma_{2},
 \epsilon_{\rm d}+\frac{\Bar{U}}{\pi}D\Big),
\end{align}
where $J(\cdot,\cdot,\cdot)$ 
is an arbitrary three-variable function.
Hence, if we change the parameters
$\Gamma_{\ell}$ and $\epsilon_{\rm d}$
as functions in $D$ as
\begin{align}
\label{eq:renormalized}
&\Gamma_{\ell}(D)
 =T_{\ell}
 \Big(\frac{T_{\rm K}}{D}\Big)^{\frac{U_{\ell}}{\pi}},\quad
 \epsilon_{\rm d}(D)
 =E_{\rm d}-\frac{\Bar{U}}{\pi}D
\end{align}
with the constants $T_{1}$, $T_{2}$, ($T_{\rm K}=T_{1}+T_{2}$)
and $E_{\rm d}$,
the average electric current $\langle I\rangle$ does not depend on $D$.
The parameters  $\Gamma_{\ell}(D)$ and $\epsilon_{\rm d}(D)$
are referred to as renormalized parameters
while the original parameters are called ``bare'' parameters, which
we denote by $\Gamma_{\ell,0}$, $\epsilon_{{\rm d},0}$ and $D_{0}$.
Here we fix the renormalized constants $T_{\ell}$ and $E_{\rm d}$ 
by the bare parameters as
\begin{align}
\label{eq:Kondo-temp}
&T_{\ell}=\Gamma_{\ell,0}
 \Big(\frac{D_{0}}{T_{\rm K}}\Big)^{\frac{U_{\ell}}{\pi}},\quad
 E_{\rm d}=\epsilon_{\rm d}+\frac{\Bar{U}}{\pi}D_{0}
\end{align}
and express all physical quantities in terms of the renormalized ones.
We shall see below that the sum $T_{\rm K}=T_{1}+T_{2}$ plays
a role of a scaling parameter similar to the Kondo temperature.

By inserting the renormalized parameters in Eqs.~\eqref{eq:renormalized}
into the average electric current 
$\langle I\rangle$ in Eq.~\eqref{eq:average-current_2}
and rearranging it with respect to the interaction parameter $U_{\ell}$,
we obtain the universal electric current
\begin{align}
\label{eq:Universal-Current_IRLM_1}
\langle I\rangle
&=-\frac{2T_{1}T_{2}}{\pi T_{\rm K}}\Tilde{j}_{-}
 \nn\\
&\quad
 +\frac{T_{1}T_{2}}{\pi^{2} T^{2}_{\rm K}}
 \sum_{\ell}U_{\ell}
 \Big[\big(T_{\Bar{\ell}}\Tilde{j}_{-}
 -T_{\ell}\Tilde{j}_{1}\big)
 \log(\Tilde{\epsilon}_{\ell}^{2}+1)
 +\big(T_{\rm K}\Tilde{\epsilon}_{\ell}-T_{\ell}
 \big(2\arctan(\Tilde{\epsilon}_{\ell})-\pi\big)\big)
 \Tilde{j}_{2}
 \Big]
 \nn\\
&\quad
 +O(U^{2}),
\end{align}
where $\Tilde{j}_{-}$ and $\Tilde{j}_{s}$ are
$j_{-}$ and $j_{s}$ with 
$\Tilde{\epsilon}_{\ell}=(E_{\rm d}-\mu_{\ell})/T_{\rm K}$
in place of $\epsilon_{\ell}$, respectively.
Thus the average electric current $\langle I\rangle$,
which was originally described by the bare parameters 
$\Gamma_{\ell, 0}$, $\epsilon_{{\rm d}, 0}$, $D_{0}$ and $U_{\ell}$,
is now characterized by the parameters 
$T_{\ell}$, $E_{\rm d}$ and $U_{\ell}$.
As a result, the linear and the logarithmic divergences
of the average electric current
are absorbed into the parameters $T_{\ell}$ and $E_{\rm d}$.

Let us consider the current-voltage ($I$-$V$) characteristics
of the universal electric current.
We put $E_{\rm d}=0$ and
consider the cases $\mu_{1}=-\mu_{2}=V/2$ with the bias voltage $V$.
Then we have
\begin{align}
\label{eq:Universal-Current_IRLM_2}
\frac{T_{\rm K}}{4T_{1}T_{2}}\langle I\rangle
&=\frac{1}{\pi}
 \arctan\Big(\frac{V}{2T_{\rm K}}\Big)
 \nn\\
&\quad
 -\frac{1}{2\pi^{2}}
 \Big[(\Bar{U}-\delta)\arctan\Big(\frac{V}{2T_{\rm K}}\Big)
 -(\Bar{U}+\delta)\frac{V/(2T_{\rm K})}{V^{2}/(2T_{\rm K})^{2}+1}\Big]
 \log\Big(\Big(\frac{V}{2T_{\rm K}}\Big)^{2}\!\!+1\Big)
 \nn\\
&\quad
 +O(U^{2}),
\end{align}
where $\delta$ is the asymmetry parameter defined by
\begin{align}
\label{eq:asymmetric}
 \delta=\frac{(U_{1}-U_{2})(T_{1}-T_{2})}{2T_{\rm K}}.
\end{align}
We note that the parameters $T_{1}$ and $T_{2}$ depend
on $U_{1}$ and $U_{2}$ through Eqs.~\eqref{eq:Kondo-temp}.

We find from Eq.~\eqref{eq:Universal-Current_IRLM_2}
that the bias voltage $V$ is scaled by the parameter $T_{\rm K}$.
In the case of symmetric interactions $U_{1}=U_{2}$, we have $\delta=0$.
Hence, after taking appropriate scaling factors 
for the electric current $\langle I\rangle$ and the bias voltage $V$, 
the $I$-$V$ curve is independent of the parameters $T_{1}$ and $T_{2}$.
In the asymmetric cases with $\Gamma_{1,0}\neq\Gamma_{2,0}$ and
$U_{1}\neq U_{2}$, on the other hand, 
the parameters $T_{1}$ and $T_{2}$ play a nontrivial role
since, even if we rescale the electric current, it still depends on 
$T_{1}$ and $T_{2}$ through the asymmetry parameter $\delta$.

\begin{figure}[t]
\includegraphics[width=250pt,clip]{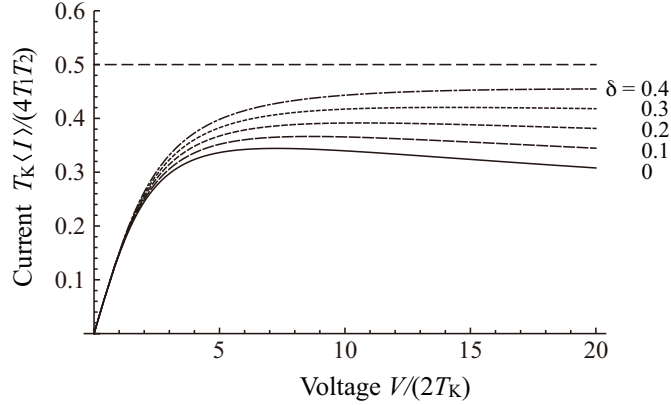}
\caption{$I$-$V$ characteristics of the rescaled universal electric current
for $\Bar{U}=0.5$ with $\delta=0, 0.1, 0.2, 0.3, 0.4$.}
\label{fig:I-V}
\end{figure}

We deal with the parameters $T_{1}$ and $T_{2}$ 
in the first order of $U_{\ell}$.
By solving the equation for $T_{\rm K}$,
which is obtained by the first equation in Eqs.~\eqref{eq:Kondo-temp}, 
the parameter $T_{\ell}$ is expanded in the first order of $U_{\ell}$ as
\begin{align}
\label{eq:Kondo-temp_2}
&T_{\ell}=\Gamma_{\ell,0}\Big(1+
 \frac{U_{\ell}}{\pi}\log\Big(\frac{D_{0}}{\Gamma_{0}}\Big)\Big)
 +O(U^{2}).
\end{align}
We remark that, for the consistency with the expansion,
the ratio $D_{0}/\Gamma_{0}$ should be restricted to the region
$D_{0}/\Gamma_{0}\ll\e^{\pi/U_{\ell}}$.
Then the asymmetry parameter $\delta$ 
in Eq.~\eqref{eq:asymmetric} is expanded as
\begin{align}
\delta
&=\frac{(U_{1}-U_{2})(\Gamma_{1,0}-\Gamma_{2,0})}{2\Gamma_{0}}+O(U^{2}).
\end{align}
By a physical intuition, we expect $\delta\geq 0$
since the case $U_{1}>U_{2}$ should correspond to the case $\Gamma_{1,0}>\Gamma_{2,0}$.
By expressing the parameters as 
$\Gamma_{1/2,0}=\Gamma_{0}(1\pm\gamma)/2$ with $0\leq\gamma<1$ and
$U_{1/2}=\Bar{U}(1\pm\gamma^{\prime})$ with $0\leq\gamma^{\prime}\leq 1$
in the cases with $\Gamma_{1,0}\geq\Gamma_{2,0}>0$
and $U_{1}\geq U_{2}$, we have $\delta=\Bar{U}\gamma\gamma^{\prime}+O(U^{2})$.
Hence the asymmetry parameter $\delta$
takes a value in the range $0\leq\delta<\Bar{U}+O(U^{2})$.
The $I$-$V$ curve of the universal electric current for $\Bar{U}=0.5$
and $0\leq\delta<0.5$ is depicted in Fig.~\ref{fig:I-V}.

We observe the suppression of the electric current
for large bias voltages $V\gg T_{\rm K}$,
that is, {\it negative differential conductance}.
This is because the formation of the two-body bound states
promotes the reflection of electrons at the QD,
as is illustrated in Fig.~\ref{fig:IRLM-asym2-2BS}, and
the logarithmic term in 
Eq.~\eqref{eq:Universal-Current_IRLM_2} decreases the electric current.
In the first order of $U_{\ell}$, 
the negative differential conductance shows the power-law behavior 
$\langle I\rangle\propto (V/T_{\rm K})^{-(\Bar{U}-\delta)/\pi}$
where the asymmetry parameter $\delta$ appears.
This means that the suppressed electric current is restored by
the asymmetry of the system parameters.
In the case $\Gamma_{1,0}=\Gamma_{2,0}$, 
we have $\delta=0$ in the first order of $U_{\ell}$
and the universal electric current 
in Eq.~\eqref{eq:Universal-Current_IRLM_2} depends
only on the average interaction $\Bar{U}$.
This is consistent with the Callan-Symanzik equation:
by changing the variables and
setting $\Gamma_{1}=\Gamma_{2}=\Gamma/2$, 
the Callan-Symanzik equation~\eqref{eq:Callan-Symanzik-eq} 
is reduced to
\begin{align}
\label{eq:Callan-Symanzik-eq_2}
&\Big(D\frac{\partial}{\partial D}
 +\beta_{\Gamma}\frac{\partial}{\partial\Gamma}
 +\beta_{\epsilon_{\rm d}}\frac{\partial}{\partial\epsilon_{\rm d}}
 \Big)
 \langle I\rangle=0
\end{align}
with the beta function
$\beta_{\Gamma}=-\Bar{U}\Gamma/\pi+O(U^{2})$.
The general solution is given in the form
$\langle I\rangle=J(D^{\frac{\Bar{U}}{\pi}}\Gamma,
\epsilon_{\rm d}+\Bar{U}D/\pi)$
with an arbitrary two-variable function $J(\cdot,\cdot)$.
Then the negative differential conductance appears for $\Bar{U}>0$, 
which is essentially the same as the previous results
in the case of symmetric interactions
$U_{1}=U_{2}$~\cite{Doyon_07PRL,Boulat-Saleur-Schmitteckert_08PRL,%
Nishino-Imamura-Hatano_11PRB}.

Let us compare the present results with 
the renormalization-group (RG) results in Refs.~\onlinecite{%
Karrasch-Andergassen-Pletyukhov_10EPL} and \onlinecite{%
Andergassen-Pletyukhov-Schuricht-Schoeller-Borda_11PRB};
the RG flow equations for the level width $\Gamma$ were obtained 
in the second order of $U_{\ell}$ in Ref.~\onlinecite{%
Andergassen-Pletyukhov-Schuricht-Schoeller-Borda_11PRB}.
Although general local dispersion relations of the leads
were adopted in Refs.~\onlinecite{%
Karrasch-Andergassen-Pletyukhov_10EPL} and \onlinecite{%
Andergassen-Pletyukhov-Schuricht-Schoeller-Borda_11PRB}, 
the details of the frequency dependence of the density of states
of the leads did not play any role in their analysis.
Indeed, the linear divergence in our average electric current
$\langle I\rangle$ due to the linearized dispersion relations
is removed by the RG technique with the Callan-Symanzik equation
as we have described above.
Their definition of the parameters $T_{1}$ and $T_{2}$,
which was approximately derived from the RG flow
equations for the level width $\Gamma$,
is equivalent to ours in Eq.~\eqref{eq:Kondo-temp}
in the first order of $U_{\ell}$.

We can confirm that the universal electric current 
in Eq.~\eqref{eq:Universal-Current_IRLM_2} is consistent with that of 
the RG results~\cite{Karrasch-Andergassen-Pletyukhov_10EPL,
Andergassen-Pletyukhov-Schuricht-Schoeller-Borda_11PRB}
in the first order of $U_{\ell}$.
By using the renormalized band width $\Gamma_{\ell}(D)$ 
in Eq.~\eqref{eq:renormalized}, we have
\begin{align}
&\frac{\Gamma_{1}(D)\Gamma_{2}(D)}{\Gamma(D)}
 \arctan\Big(\frac{V}{2\Gamma(D)}\Big)
 \nn\\
&=\frac{T_{1}T_{2}}{T_{\rm K}}
 \Big[\arctan\Big(\frac{V}{2T_{\rm K}}\Big)
 \nn\\
&\quad -\!\frac{1}{2\pi}\Big[
 (\Bar{U}\!-\!\delta)\arctan\Big(\frac{V}{2T_{\rm K}}\Big)
 -(\Bar{U}\!+\!\delta)\frac{V/(2T_{\rm K})}{V^{2}/(2T_{\rm K})^{2}\!+\!1}
 \Big]
 \log\Big(\frac{D^{2}}{T^{2}_{\rm K}}\Big)\Big]
 +O(U^{2}).
\end{align}
Then, by putting $\Tilde{\Gamma}_{\ell}=\Gamma_{\ell}(|V/2+\i T_{\rm K}|)$
and $\Tilde{\Gamma}=\Tilde{\Gamma}_{1}+\Tilde{\Gamma}_{2}$, 
the universal electric current in Eq.~\eqref{eq:Universal-Current_IRLM_2}
is expressed by
\begin{align}
\label{eq:Universal-Current_IRLM_3}
 \langle I\rangle
 =\frac{4\Tilde{\Gamma}_{1}\Tilde{\Gamma}_{2}}{\pi\Tilde{\Gamma}}
 \arctan\Big(\frac{V}{2\Tilde{\Gamma}}\Big)+O(U^{2}),
\end{align}
which is in the same form as the noninteracting cases.
The expression in Eq.~\eqref{eq:Universal-Current_IRLM_3}
agrees with that obtained in the RG results (see Appendix~\ref{sec:RG-results}).

We remark that, although the expression of the universal electric current
in Refs.~\onlinecite{Karrasch-Andergassen-Pletyukhov_10EPL} and
\onlinecite{Andergassen-Pletyukhov-Schuricht-Schoeller-Borda_11PRB} agrees with ours
in Eq.~\eqref{eq:Universal-Current_IRLM_3}, their treatment of
the renormalized parameters $T_{1}$ and $T_{2}$ included in the universal electric current 
is different from ours;
they used the higher-order terms of $U_{\ell}$ 
in the defining relations of $T_{1}$ and $T_{2}$ in Eq.~\eqref{eq:Kondo-temp}
while we have treated them in the first order of $U_{\ell}$ 
as is given Eq.~\eqref{eq:Kondo-temp_2}.
As a result, even in the case $\Gamma_{1,0}=\Gamma_{2,0}$
of symmetric lead-dot couplings,
they observed the restoration of the suppressed electric current.
We have shown that, for the asymmetry parameter $\delta$
treated in the first order of $U_{\ell}$,
the restoration due to the asymmetric interactions
does not occur at $\Gamma_{1,0}=\Gamma_{2,0}$.
In other words, in this case, there should be no restoration 
of the suppressed electric current for small $U_{\ell}$, 
which seems to differ from the results of Refs.~\onlinecite{%
Karrasch-Andergassen-Pletyukhov_10EPL} and \onlinecite{%
Andergassen-Pletyukhov-Schuricht-Schoeller-Borda_11PRB}.

\section{Concluding remarks}
\label{sec:concluding-remarks}

We have studied the average electric current for the open QD systems 
described by the two-lead IRLM in the asymmetric settings.
By using the extension of the Landauer formula
with the many-electron scattering eigenstates,
we have calculated the average electric current for the systems
under finite bias voltages.
The calculation is in the first order of the interaction parameters, 
but otherwise we have not employed any approximations.
The calculation itself has been considerably
simplified compared to the previous one~\cite{%
Nishino-Imamura-Hatano_09PRL,Nishino-Imamura-Hatano_11PRB} 
treating the symmetric cases with the even-odd transformation.

Through the renormalization-group technique with the Callan-Symanzik equation,
we have obtained the universal electric current characterized 
by the scaling parameter $T_{\rm K}$ and the asymmetry parameter $\delta$.
The Coulomb interactions around the QD give rise to
the negative differential conductance through
the formation of the two-body bound states,
which is a new point of view clarified in our analysis.
Through the investigation of the asymmetry parameter $\delta$,
we have confirmed in the first order of $U_{1}$ and $U_{2}$ that
the suppressed electric current is restored in the asymmetric cases 
satisfying both $\Gamma_{1,0}\neq\Gamma_{2,0}$ and $U_{1}\neq U_{2}$.

The analytic form of the universal electric current has enabled
us to compare it with those of other approaches correctly.
Our universal electric current has the same functional form as that 
obtained with the RG approach~\cite{%
Karrasch-Andergassen-Pletyukhov_10EPL,
Andergassen-Pletyukhov-Schuricht-Schoeller-Borda_11PRB}; 
it  also reproduces previous results~\cite{%
Borda-Vladar-Zawadowski_07PRB,Doyon_07PRL,%
Boulat-Saleur-Schmitteckert_08PRL,Golub_07PRB,%
Nishino-Imamura-Hatano_11PRB} in the symmetric cases. 
However, for $U_{1}\neq U_{2}$ and $\Gamma_{1,0}=\Gamma_{2,0}$, 
our results indicate that there is no restoration of the suppressed electric current
to first order in $U_{1}$ and $U_{2}$, 
while Refs.~\onlinecite{Karrasch-Andergassen-Pletyukhov_10EPL} and
\onlinecite{Andergassen-Pletyukhov-Schuricht-Schoeller-Borda_11PRB} 
indicate that there is. 
This restoration may 
result from higher-order terms in $U_{1}$ and $U_{2}$.
To verify its validity analytically, we need a consistent treatment 
of these higher-order terms, which can be done with 
our extension of the Landauer formula.

The key element for the practical calculations in the extension
of the Landauer formula is the explicit form of many-electron scattering eigenstates.
The present calculation in the first order in the interaction parameters
can be extended to higher orders by using the exact $N$-electron 
scattering eigenstates that we have already obtained.
We expect that such calculation should be applied to other physical quantities
of the open QD systems such as the dot occupancy.

\begin{acknowledgments}
This work was supported by JSPS KAKENHI 
Grant Numbers 22340110, 23740301, 26400409
\end{acknowledgments}

\appendix

\section{Results of the RG flow equation for $\Gamma$}
\label{sec:RG-results}

In our settings, the universal electric current obtained 
in Refs.~\onlinecite{Karrasch-Andergassen-Pletyukhov_10EPL} and
\onlinecite{Andergassen-Pletyukhov-Schuricht-Schoeller-Borda_11PRB} 
is expressed as
\begin{align}
\label{eq:Universal-Current_IRLM_RG}
 \langle I\rangle_{\rm RG}
&=\frac{2\Hat{\Gamma}_{1}\Hat{\Gamma}_{2}}{\pi\Hat{\Gamma}}
  \Big[\arctan\Big(\frac{\mu_{1}\!-\!\epsilon_{\rm d}}{\Hat{\Gamma}}\Big)
  \!-\!\arctan\Big(\frac{\mu_{2}\!-\!\epsilon_{\rm d}}{\Hat{\Gamma}}\Big)\Big],
\end{align}
where $\Hat{\Gamma}_{\ell}$ is the solution of the RG flow equations
for the level width of the QD and $\Hat{\Gamma}=\Hat{\Gamma}_{1}+\Hat{\Gamma}_{2}$. 
It was approximately determined by the self-consistent equation
\begin{align}
 \Hat{\Gamma}_{\ell}\simeq \Gamma_{\ell,0}
 \Big(\frac{D_{0}}
 {|\Hat{\Gamma}-\i(\mu_{\ell}-\epsilon_{\rm d})|}\Big)^{g_{\ell}}
\end{align}
with $g_{\ell}=U_{\ell}/\pi+O(U^{2})$.
In the first order of $U_{\ell}$, the approximate solution is given by
\begin{align}
 \Hat{\Gamma}_{\ell}
&\simeq T_{\ell}
 \Big(\frac{T_{\rm K}}
 {|\Hat{\Gamma}-\i(\mu_{\ell}-\epsilon_{\rm d})|}\Big)^{\frac{U_{\ell}}{\pi}}+O(U^{2})
 \nn\\
&=T_{\ell}
 \Big(\frac{T_{\rm K}}
 {|T_{\rm K}-\i(\mu_{\ell}-\epsilon_{\rm d})|}\Big)^{\frac{U_{\ell}}{\pi}}+O(U^{2})
 \nn\\
&=\Gamma_{\ell}(|(\mu_{\ell}-\epsilon_{\rm d})+\i T_{\rm K}|)+O(U^{2}),
\end{align}
where $\Gamma_{\ell}(D)$ is the renormalized level width defined in Eq.~\eqref{eq:renormalized}.
By inserting this into Eq.~\eqref{eq:Universal-Current_IRLM_RG}
and setting $\mu_{1/2}=\pm V/2$ and $\epsilon_{\rm d}=0$,
the expression in Eq.~\eqref{eq:Universal-Current_IRLM_RG}
agrees with Eq.~\eqref{eq:Universal-Current_IRLM_3}.

\bibliographystyle{apsrev4-1}

\begin{thebibliography}{42}%
\makeatletter
\providecommand \@ifxundefined [1]{%
 \@ifx{#1\undefined}
}%
\providecommand \@ifnum [1]{%
 \ifnum #1\expandafter \@firstoftwo
 \else \expandafter \@secondoftwo
 \fi
}%
\providecommand \@ifx [1]{%
 \ifx #1\expandafter \@firstoftwo
 \else \expandafter \@secondoftwo
 \fi
}%
\providecommand \natexlab [1]{#1}%
\providecommand \enquote  [1]{``#1''}%
\providecommand \bibnamefont  [1]{#1}%
\providecommand \bibfnamefont [1]{#1}%
\providecommand \citenamefont [1]{#1}%
\providecommand \href@noop [0]{\@secondoftwo}%
\providecommand \href [0]{\begingroup \@sanitize@url \@href}%
\providecommand \@href[1]{\@@startlink{#1}\@@href}%
\providecommand \@@href[1]{\endgroup#1\@@endlink}%
\providecommand \@sanitize@url [0]{\catcode `\\12\catcode `\$12\catcode
  `\&12\catcode `\#12\catcode `\^12\catcode `\_12\catcode `\%12\relax}%
\providecommand \@@startlink[1]{}%
\providecommand \@@endlink[0]{}%
\providecommand \url  [0]{\begingroup\@sanitize@url \@url }%
\providecommand \@url [1]{\endgroup\@href {#1}{\urlprefix }}%
\providecommand \urlprefix  [0]{URL }%
\providecommand \Eprint [0]{\href }%
\providecommand \doibase [0]{http://dx.doi.org/}%
\providecommand \selectlanguage [0]{\@gobble}%
\providecommand \bibinfo  [0]{\@secondoftwo}%
\providecommand \bibfield  [0]{\@secondoftwo}%
\providecommand \translation [1]{[#1]}%
\providecommand \BibitemOpen [0]{}%
\providecommand \bibitemStop [0]{}%
\providecommand \bibitemNoStop [0]{.\EOS\space}%
\providecommand \EOS [0]{\spacefactor3000\relax}%
\providecommand \BibitemShut  [1]{\csname bibitem#1\endcsname}%
\let\auto@bib@innerbib\@empty
\bibitem [{\citenamefont {{D. Goldhaber-Gordon, Hadas Shtrikman, D. Mahalu,
  David Abusch-Magder, U. Meirav, and M. A.
  Kastner}}(1998)}]{GoldhaberGordon-Shtrikman-Mahalu-AbuschMagder-Meirav-Kastner_98Nature}%
  \BibitemOpen
  \bibfield  {author} {\bibinfo {author} {\bibnamefont {{D. Goldhaber-Gordon,
  Hadas Shtrikman, D. Mahalu, David Abusch-Magder, U. Meirav, and M. A.
  Kastner}}},\ }\href@noop {} {\bibfield  {journal} {\bibinfo  {journal}
  {Nature (London)}\ }\textbf {\bibinfo {volume} {391}},\ \bibinfo {pages}
  {156} (\bibinfo {year} {1998})}\BibitemShut {NoStop}%
\bibitem [{\citenamefont {Cronenwett}\ \emph {et~al.}(1998)\citenamefont
  {Cronenwett}, \citenamefont {Oosterkamp},\ and\ \citenamefont
  {Kouwenhoven}}]{Cronenwett-Oosterkamp-Kouwenhoven_98Science}%
  \BibitemOpen
  \bibfield  {author} {\bibinfo {author} {\bibfnamefont {S.~M.}\ \bibnamefont
  {Cronenwett}}, \bibinfo {author} {\bibfnamefont {T.~H.}\ \bibnamefont
  {Oosterkamp}}, \ and\ \bibinfo {author} {\bibfnamefont {L.~P.}\ \bibnamefont
  {Kouwenhoven}},\ }\href@noop {} {\bibfield  {journal} {\bibinfo  {journal}
  {Science}\ }\textbf {\bibinfo {volume} {281}},\ \bibinfo {pages} {540}
  (\bibinfo {year} {1998})}\BibitemShut {NoStop}%
\bibitem [{\citenamefont {{W. G. van der Wiel, S. De Franceschi, T. Fujisawa,
  J. M. Elzerman, S. Tarucha, and L. P.
  Kouwenhoven}}(2000)}]{Wiel-Franceschi-Fujisawa-Elzerman-Tarucha-Kouwenhoven_00Science}%
  \BibitemOpen
  \bibfield  {author} {\bibinfo {author} {\bibnamefont {{W. G. van der Wiel, S.
  De Franceschi, T. Fujisawa, J. M. Elzerman, S. Tarucha, and L. P.
  Kouwenhoven}}},\ }\href@noop {} {\bibfield  {journal} {\bibinfo  {journal}
  {Science}\ }\textbf {\bibinfo {volume} {289}},\ \bibinfo {pages} {2105}
  (\bibinfo {year} {2000})}\BibitemShut {NoStop}%
\bibitem [{\citenamefont {{Andrey V. Kretinin, Hadas Shtrikman, David
  Goldhaber-Gordon, Markus Hanl, Andreas Weichselbaum, Jan von Delft, Theo
  Costi, and Diana
  Mahalu}}(2011)}]{Kretinin-Shtrikman-GoldhaberGordon-Hanl-Weichselbaum-vonDelft-Costi-Mahalu_11PRB}%
  \BibitemOpen
  \bibfield  {author} {\bibinfo {author} {\bibnamefont {{Andrey V. Kretinin,
  Hadas Shtrikman, David Goldhaber-Gordon, Markus Hanl, Andreas Weichselbaum,
  Jan von Delft, Theo Costi, and Diana Mahalu}}},\ }\href@noop {} {\bibfield
  {journal} {\bibinfo  {journal} {Phys.\ Rev.\ B}\ }\textbf {\bibinfo {volume}
  {84}},\ \bibinfo {pages} {245316} (\bibinfo {year} {2011})}\BibitemShut
  {NoStop}%
\bibitem [{\citenamefont {{A. Yacoby, M. Heiblum, D. Mahalu, and Hadas
  Shtrikman}}(1995)}]{Yacoby-Heiblum-Mahalu-Shtrikman_95PRL}%
  \BibitemOpen
  \bibfield  {author} {\bibinfo {author} {\bibnamefont {{A. Yacoby, M. Heiblum,
  D. Mahalu, and Hadas Shtrikman}}},\ }\href@noop {} {\bibfield  {journal}
  {\bibinfo  {journal} {Phys.\ Rev.\ Lett.}\ }\textbf {\bibinfo {volume}
  {74}},\ \bibinfo {pages} {4047} (\bibinfo {year} {1995})}\BibitemShut
  {NoStop}%
\bibitem [{\citenamefont {{R. Schuster, E. Buks, M. Heiblum, D. Mahalu, V.
  Umansky, and Hadas
  Shtrikman}}(1997)}]{Schuster-Buks-Heiblum-Mahalu-Umansky-Shtrikman_97Nature}%
  \BibitemOpen
  \bibfield  {author} {\bibinfo {author} {\bibnamefont {{R. Schuster, E. Buks,
  M. Heiblum, D. Mahalu, V. Umansky, and Hadas Shtrikman}}},\ }\href@noop {}
  {\bibfield  {journal} {\bibinfo  {journal} {Nature}\ }\textbf {\bibinfo
  {volume} {385}},\ \bibinfo {pages} {417} (\bibinfo {year}
  {1997})}\BibitemShut {NoStop}%
\bibitem [{\citenamefont {Datta}(1995)}]{Datta}%
  \BibitemOpen
  \bibfield  {author} {\bibinfo {author} {\bibfnamefont {S.}~\bibnamefont
  {Datta}},\ }\href@noop {} {\emph {\bibinfo {title} {Electronic Transport in
  Mesoscopic Systems}}}\ (\bibinfo  {publisher} {Cambridge University Press,
  Cambridge},\ \bibinfo {year} {1995})\BibitemShut {NoStop}%
\bibitem [{\citenamefont {Imry}(2002)}]{Imry}%
  \BibitemOpen
  \bibfield  {author} {\bibinfo {author} {\bibfnamefont {Y.}~\bibnamefont
  {Imry}},\ }\href@noop {} {\emph {\bibinfo {title} {Introduction to Mesoscopic
  Physics}}},\ \bibinfo {edition} {2nd}\ ed.\ (\bibinfo  {publisher} {Oxford
  Univ. Pr., Oxford},\ \bibinfo {year} {2002})\BibitemShut {NoStop}%
\bibitem [{\citenamefont {Yurke}\ and\ \citenamefont
  {Kochanski}(1990)}]{Yurke-Kochanski_90PRB}%
  \BibitemOpen
  \bibfield  {author} {\bibinfo {author} {\bibfnamefont {B.}~\bibnamefont
  {Yurke}}\ and\ \bibinfo {author} {\bibfnamefont {G.~P.}\ \bibnamefont
  {Kochanski}},\ }\href@noop {} {\bibfield  {journal} {\bibinfo  {journal}
  {Phys.\ Rev.\ B}\ }\textbf {\bibinfo {volume} {41}},\ \bibinfo {pages} {8184}
  (\bibinfo {year} {1990})}\BibitemShut {NoStop}%
\bibitem [{\citenamefont {{M. B\"{u}ttiker}}(1992)}]{Buttiker_92PRB}%
  \BibitemOpen
  \bibfield  {author} {\bibinfo {author} {\bibnamefont {{M. B\"{u}ttiker}}},\
  }\href@noop {} {\bibfield  {journal} {\bibinfo  {journal} {Phys.\ Rev.\ B}\
  }\textbf {\bibinfo {volume} {46}},\ \bibinfo {pages} {12485} (\bibinfo {year}
  {1992})}\BibitemShut {NoStop}%
\bibitem [{\citenamefont {Blanter}\ and\ \citenamefont {{M.
  B\"{u}ttiker}}(2000)}]{Blanter-Buttiker_00PR}%
  \BibitemOpen
  \bibfield  {author} {\bibinfo {author} {\bibfnamefont {Y.~M.}\ \bibnamefont
  {Blanter}}\ and\ \bibinfo {author} {\bibnamefont {{M. B\"{u}ttiker}}},\
  }\href@noop {} {\bibfield  {journal} {\bibinfo  {journal} {Phys.\ Rep.}\
  }\textbf {\bibinfo {volume} {336}},\ \bibinfo {pages} {1} (\bibinfo {year}
  {2000})}\BibitemShut {NoStop}%
\bibitem [{\citenamefont {{C. Caroli, R. Combescot, P. Nozieres, and D.
  Saint-James}}(1971)}]{Caroli-Combescot-Nozieres-SaintJames_71JPC}%
  \BibitemOpen
  \bibfield  {author} {\bibinfo {author} {\bibnamefont {{C. Caroli, R.
  Combescot, P. Nozieres, and D. Saint-James}}},\ }\href@noop {} {\bibfield
  {journal} {\bibinfo  {journal} {J.\ Phys.\ C: Solid State Phys.}\ }\textbf
  {\bibinfo {volume} {4}},\ \bibinfo {pages} {916} (\bibinfo {year}
  {1971})}\BibitemShut {NoStop}%
\bibitem [{\citenamefont {Meir}\ \emph {et~al.}(1991)\citenamefont {Meir},
  \citenamefont {Wingreen},\ and\ \citenamefont
  {Lee}}]{Meir-Wingreen-Lee_91PRL}%
  \BibitemOpen
  \bibfield  {author} {\bibinfo {author} {\bibfnamefont {Y.}~\bibnamefont
  {Meir}}, \bibinfo {author} {\bibfnamefont {N.~S.}\ \bibnamefont {Wingreen}},
  \ and\ \bibinfo {author} {\bibfnamefont {P.~A.}\ \bibnamefont {Lee}},\
  }\href@noop {} {\bibfield  {journal} {\bibinfo  {journal} {Phys.\ Rev.\
  Lett.}\ }\textbf {\bibinfo {volume} {66}},\ \bibinfo {pages} {3048} (\bibinfo
  {year} {1991})}\BibitemShut {NoStop}%
\bibitem [{\citenamefont {Meir}\ and\ \citenamefont
  {Wingreen}(1992)}]{Meir-Wingreen_92PRL}%
  \BibitemOpen
  \bibfield  {author} {\bibinfo {author} {\bibfnamefont {Y.}~\bibnamefont
  {Meir}}\ and\ \bibinfo {author} {\bibfnamefont {N.~S.}\ \bibnamefont
  {Wingreen}},\ }\href@noop {} {\bibfield  {journal} {\bibinfo  {journal}
  {Phys.\ Rev.\ Lett.}\ }\textbf {\bibinfo {volume} {68}},\ \bibinfo {pages}
  {2512} (\bibinfo {year} {1992})}\BibitemShut {NoStop}%
\bibitem [{\citenamefont {Hershfield}\ \emph {et~al.}(1992)\citenamefont
  {Hershfield}, \citenamefont {Davies},\ and\ \citenamefont
  {Wilkins}}]{Hershfield-Davies-Wilkins_92PRB}%
  \BibitemOpen
  \bibfield  {author} {\bibinfo {author} {\bibfnamefont {S.}~\bibnamefont
  {Hershfield}}, \bibinfo {author} {\bibfnamefont {J.~H.}\ \bibnamefont
  {Davies}}, \ and\ \bibinfo {author} {\bibfnamefont {J.~W.}\ \bibnamefont
  {Wilkins}},\ }\href@noop {} {\bibfield  {journal} {\bibinfo  {journal}
  {Phys.\ Rev.\ B}\ }\textbf {\bibinfo {volume} {46}},\ \bibinfo {pages} {7046}
  (\bibinfo {year} {1992})}\BibitemShut {NoStop}%
\bibitem [{\citenamefont {Haug}\ and\ \citenamefont
  {Jauho}(2007)}]{Haug-Jauho}%
  \BibitemOpen
  \bibfield  {author} {\bibinfo {author} {\bibfnamefont {H.}~\bibnamefont
  {Haug}}\ and\ \bibinfo {author} {\bibfnamefont {A.~P.}\ \bibnamefont
  {Jauho}},\ }\href@noop {} {\emph {\bibinfo {title} {Quantum Kinetics in
  Transport and Optics of Semiconductors}}},\ \bibinfo {edition} {2nd}\ ed.\
  (\bibinfo  {publisher} {Springer, Berlin},\ \bibinfo {year}
  {2007})\BibitemShut {NoStop}%
\bibitem [{\citenamefont {Ng}\ and\ \citenamefont {Lee}(1988)}]{Ng-Lee_88PRL}%
  \BibitemOpen
  \bibfield  {author} {\bibinfo {author} {\bibfnamefont {T.~K.}\ \bibnamefont
  {Ng}}\ and\ \bibinfo {author} {\bibfnamefont {P.~A.}\ \bibnamefont {Lee}},\
  }\href@noop {} {\bibfield  {journal} {\bibinfo  {journal} {Phys.\ Rev.\
  Lett.}\ }\textbf {\bibinfo {volume} {61}},\ \bibinfo {pages} {1768} (\bibinfo
  {year} {1988})}\BibitemShut {NoStop}%
\bibitem [{\citenamefont {Wingreen}\ and\ \citenamefont
  {Meir}(1994)}]{Wingreen-Meir_94PRB}%
  \BibitemOpen
  \bibfield  {author} {\bibinfo {author} {\bibfnamefont {N.~S.}\ \bibnamefont
  {Wingreen}}\ and\ \bibinfo {author} {\bibfnamefont {Y.}~\bibnamefont
  {Meir}},\ }\href@noop {} {\bibfield  {journal} {\bibinfo  {journal} {Phys.\
  Rev.\ B}\ }\textbf {\bibinfo {volume} {49}},\ \bibinfo {pages} {11040}
  (\bibinfo {year} {1994})}\BibitemShut {NoStop}%
\bibitem [{\citenamefont {Nishino}\ \emph {et~al.}(2009)\citenamefont
  {Nishino}, \citenamefont {Imamura},\ and\ \citenamefont
  {Hatano}}]{Nishino-Imamura-Hatano_09PRL}%
  \BibitemOpen
  \bibfield  {author} {\bibinfo {author} {\bibfnamefont {A.}~\bibnamefont
  {Nishino}}, \bibinfo {author} {\bibfnamefont {T.}~\bibnamefont {Imamura}}, \
  and\ \bibinfo {author} {\bibfnamefont {N.}~\bibnamefont {Hatano}},\
  }\href@noop {} {\bibfield  {journal} {\bibinfo  {journal} {Phys.\ Rev.\
  Lett.}\ }\textbf {\bibinfo {volume} {102}},\ \bibinfo {pages} {146803}
  (\bibinfo {year} {2009})}\BibitemShut {NoStop}%
\bibitem [{\citenamefont {Nishino}\ \emph {et~al.}(2011)\citenamefont
  {Nishino}, \citenamefont {Imamura},\ and\ \citenamefont
  {Hatano}}]{Nishino-Imamura-Hatano_11PRB}%
  \BibitemOpen
  \bibfield  {author} {\bibinfo {author} {\bibfnamefont {A.}~\bibnamefont
  {Nishino}}, \bibinfo {author} {\bibfnamefont {T.}~\bibnamefont {Imamura}}, \
  and\ \bibinfo {author} {\bibfnamefont {N.}~\bibnamefont {Hatano}},\
  }\href@noop {} {\bibfield  {journal} {\bibinfo  {journal} {Phys.\ Rev.\ B}\
  }\textbf {\bibinfo {volume} {83}},\ \bibinfo {pages} {035306} (\bibinfo
  {year} {2011})}\BibitemShut {NoStop}%
\bibitem [{\citenamefont {Filyov}\ and\ \citenamefont
  {Wiegmann}(1980)}]{Filyov-Wiegmann_80PLA}%
  \BibitemOpen
  \bibfield  {author} {\bibinfo {author} {\bibfnamefont {V.~M.}\ \bibnamefont
  {Filyov}}\ and\ \bibinfo {author} {\bibfnamefont {P.~B.}\ \bibnamefont
  {Wiegmann}},\ }\href@noop {} {\bibfield  {journal} {\bibinfo  {journal}
  {Phys.\ Lett.\ A}\ }\textbf {\bibinfo {volume} {76}},\ \bibinfo {pages} {283}
  (\bibinfo {year} {1980})}\BibitemShut {NoStop}%
\bibitem [{\citenamefont {Mehta}\ and\ \citenamefont
  {Andrei}(2006)}]{Mehta-Andrei_06PRL}%
  \BibitemOpen
  \bibfield  {author} {\bibinfo {author} {\bibfnamefont {P.}~\bibnamefont
  {Mehta}}\ and\ \bibinfo {author} {\bibfnamefont {N.}~\bibnamefont {Andrei}},\
  }\href@noop {} {\bibfield  {journal} {\bibinfo  {journal} {Phys.\ Rev.\
  Lett.}\ }\textbf {\bibinfo {volume} {96}},\ \bibinfo {pages} {216802}
  (\bibinfo {year} {2006})}\BibitemShut {NoStop}%
\bibitem [{\citenamefont {Borda}\ \emph {et~al.}(2007)\citenamefont {Borda},
  \citenamefont {Vlad\'ar},\ and\ \citenamefont
  {Zawadowski}}]{Borda-Vladar-Zawadowski_07PRB}%
  \BibitemOpen
  \bibfield  {author} {\bibinfo {author} {\bibfnamefont {L.}~\bibnamefont
  {Borda}}, \bibinfo {author} {\bibfnamefont {K.}~\bibnamefont {Vlad\'ar}}, \
  and\ \bibinfo {author} {\bibfnamefont {A.}~\bibnamefont {Zawadowski}},\
  }\href@noop {} {\bibfield  {journal} {\bibinfo  {journal} {Phys.\ Rev.\ B}\
  }\textbf {\bibinfo {volume} {75}},\ \bibinfo {pages} {125107} (\bibinfo
  {year} {2007})}\BibitemShut {NoStop}%
\bibitem [{\citenamefont {Doyon}(2007)}]{Doyon_07PRL}%
  \BibitemOpen
  \bibfield  {author} {\bibinfo {author} {\bibfnamefont {B.}~\bibnamefont
  {Doyon}},\ }\href@noop {} {\bibfield  {journal} {\bibinfo  {journal} {Phys.\
  Rev.\ Lett.}\ }\textbf {\bibinfo {volume} {99}},\ \bibinfo {pages} {076806}
  (\bibinfo {year} {2007})}\BibitemShut {NoStop}%
\bibitem [{\citenamefont {Boulat}\ \emph {et~al.}(2008)\citenamefont {Boulat},
  \citenamefont {Saleur},\ and\ \citenamefont
  {Schmitteckert}}]{Boulat-Saleur-Schmitteckert_08PRL}%
  \BibitemOpen
  \bibfield  {author} {\bibinfo {author} {\bibfnamefont {E.}~\bibnamefont
  {Boulat}}, \bibinfo {author} {\bibfnamefont {H.}~\bibnamefont {Saleur}}, \
  and\ \bibinfo {author} {\bibfnamefont {P.}~\bibnamefont {Schmitteckert}},\
  }\href@noop {} {\bibfield  {journal} {\bibinfo  {journal} {Phys.\ Rev.\
  Lett.}\ }\textbf {\bibinfo {volume} {101}},\ \bibinfo {pages} {140601}
  (\bibinfo {year} {2008})}\BibitemShut {NoStop}%
\bibitem [{\citenamefont {Doyon}\ and\ \citenamefont
  {Andrei}(2006)}]{Doyon-Andrei_06PRB}%
  \BibitemOpen
  \bibfield  {author} {\bibinfo {author} {\bibfnamefont {B.}~\bibnamefont
  {Doyon}}\ and\ \bibinfo {author} {\bibfnamefont {N.}~\bibnamefont {Andrei}},\
  }\href@noop {} {\bibfield  {journal} {\bibinfo  {journal} {Phys.\ Rev.\ B}\
  }\textbf {\bibinfo {volume} {73}},\ \bibinfo {pages} {245326} (\bibinfo
  {year} {2006})}\BibitemShut {NoStop}%
\bibitem [{\citenamefont {Golub}(2007)}]{Golub_07PRB}%
  \BibitemOpen
  \bibfield  {author} {\bibinfo {author} {\bibfnamefont {A.}~\bibnamefont
  {Golub}},\ }\href@noop {} {\bibfield  {journal} {\bibinfo  {journal} {Phys.\
  Rev.\ B}\ }\textbf {\bibinfo {volume} {76}},\ \bibinfo {pages} {193307}
  (\bibinfo {year} {2007})}\BibitemShut {NoStop}%
\bibitem [{\citenamefont {Aharony}\ \emph {et~al.}(2000)\citenamefont
  {Aharony}, \citenamefont {Entin-Wohlman},\ and\ \citenamefont
  {Imry}}]{Aharony-EntinWohlman-Imry_00PRB}%
  \BibitemOpen
  \bibfield  {author} {\bibinfo {author} {\bibfnamefont {A.}~\bibnamefont
  {Aharony}}, \bibinfo {author} {\bibfnamefont {O.}~\bibnamefont
  {Entin-Wohlman}}, \ and\ \bibinfo {author} {\bibfnamefont {Y.}~\bibnamefont
  {Imry}},\ }\href@noop {} {\bibfield  {journal} {\bibinfo  {journal} {Phys.\
  Rev.\ B}\ }\textbf {\bibinfo {volume} {61}},\ \bibinfo {pages} {5452}
  (\bibinfo {year} {2000})}\BibitemShut {NoStop}%
\bibitem [{\citenamefont {Goorden}\ and\ \citenamefont {{M.
  B\"{u}ttiker}}(2007)}]{Goorden-Buttiker_07PRL}%
  \BibitemOpen
  \bibfield  {author} {\bibinfo {author} {\bibfnamefont {M.~C.}\ \bibnamefont
  {Goorden}}\ and\ \bibinfo {author} {\bibnamefont {{M. B\"{u}ttiker}}},\
  }\href@noop {} {\bibfield  {journal} {\bibinfo  {journal} {Phys.\ Rev.\
  Lett.}\ }\textbf {\bibinfo {volume} {99}},\ \bibinfo {pages} {146801}
  (\bibinfo {year} {2007})}\BibitemShut {NoStop}%
\bibitem [{\citenamefont {Lebedev}\ \emph {et~al.}(2008)\citenamefont
  {Lebedev}, \citenamefont {Lesovik},\ and\ \citenamefont
  {Blatter}}]{Lebedev-Lesovik-Blatter_08PRL}%
  \BibitemOpen
  \bibfield  {author} {\bibinfo {author} {\bibfnamefont {A.~V.}\ \bibnamefont
  {Lebedev}}, \bibinfo {author} {\bibfnamefont {G.~B.}\ \bibnamefont
  {Lesovik}}, \ and\ \bibinfo {author} {\bibfnamefont {G.}~\bibnamefont
  {Blatter}},\ }\href@noop {} {\bibfield  {journal} {\bibinfo  {journal}
  {Phys.\ Rev.\ Lett.}\ }\textbf {\bibinfo {volume} {100}},\ \bibinfo {pages}
  {226805} (\bibinfo {year} {2008})}\BibitemShut {NoStop}%
\bibitem [{\citenamefont {{C. Karrasch, S. Andergassen, M. Pletyukhov, D.
  Schuricht, L. Borda, V. Meden, and H.
  Schoeller}}(2010)}]{Karrasch-Andergassen-Pletyukhov_10EPL}%
  \BibitemOpen
  \bibfield  {author} {\bibinfo {author} {\bibnamefont {{C. Karrasch, S.
  Andergassen, M. Pletyukhov, D. Schuricht, L. Borda, V. Meden, and H.
  Schoeller}}},\ }\href@noop {} {\bibfield  {journal} {\bibinfo  {journal}
  {Eur.\ Phys.\ Lett.}\ }\textbf {\bibinfo {volume} {90}},\ \bibinfo {pages}
  {30003} (\bibinfo {year} {2010})}\BibitemShut {NoStop}%
\bibitem [{\citenamefont {{S. Andergassen, M. Pletyukhov, D. Schuricht, H.
  Schoeller, and L.
  Borda}}(2011)}]{Andergassen-Pletyukhov-Schuricht-Schoeller-Borda_11PRB}%
  \BibitemOpen
  \bibfield  {author} {\bibinfo {author} {\bibnamefont {{S. Andergassen, M.
  Pletyukhov, D. Schuricht, H. Schoeller, and L. Borda}}},\ }\href@noop {}
  {\bibfield  {journal} {\bibinfo  {journal} {Phys.\ Rev.\ B}\ }\textbf
  {\bibinfo {volume} {83}},\ \bibinfo {pages} {205103} (\bibinfo {year}
  {2011})}\BibitemShut {NoStop}%
\bibitem [{\citenamefont {Schuricht}()}]{Schuricht_2012}%
  \BibitemOpen
  \bibfield  {author} {\bibinfo {author} {\bibfnamefont {D.}~\bibnamefont
  {Schuricht}},\ }\href@noop {} {\ }\bibinfo {note} {{private
  communication}}\BibitemShut {NoStop}%
\bibitem [{\citenamefont {Imamura}\ \emph {et~al.}(2009)\citenamefont
  {Imamura}, \citenamefont {Nishino},\ and\ \citenamefont
  {Hatano}}]{Imamura-Nishino-Hatano_09PRB}%
  \BibitemOpen
  \bibfield  {author} {\bibinfo {author} {\bibfnamefont {T.}~\bibnamefont
  {Imamura}}, \bibinfo {author} {\bibfnamefont {A.}~\bibnamefont {Nishino}}, \
  and\ \bibinfo {author} {\bibfnamefont {N.}~\bibnamefont {Hatano}},\
  }\href@noop {} {\bibfield  {journal} {\bibinfo  {journal} {Phys.\ Rev.\ B}\
  }\textbf {\bibinfo {volume} {80}},\ \bibinfo {pages} {245323} (\bibinfo
  {year} {2009})}\BibitemShut {NoStop}%
\bibitem [{\citenamefont {Nishino}\ \emph {et~al.}(2012)\citenamefont
  {Nishino}, \citenamefont {Imamura},\ and\ \citenamefont
  {Hatano}}]{Nishino-Imamura-Hatano_12JPCS}%
  \BibitemOpen
  \bibfield  {author} {\bibinfo {author} {\bibfnamefont {A.}~\bibnamefont
  {Nishino}}, \bibinfo {author} {\bibfnamefont {T.}~\bibnamefont {Imamura}}, \
  and\ \bibinfo {author} {\bibfnamefont {N.}~\bibnamefont {Hatano}},\
  }\href@noop {} {\bibfield  {journal} {\bibinfo  {journal} {J.\ Phys.:\ Conf.\
  Ser.}\ }\textbf {\bibinfo {volume} {343}},\ \bibinfo {pages} {012087}
  (\bibinfo {year} {2012})}\BibitemShut {NoStop}%
\bibitem [{\citenamefont {Hershfield}(1993)}]{Hershfield_93PRL}%
  \BibitemOpen
  \bibfield  {author} {\bibinfo {author} {\bibfnamefont {S.}~\bibnamefont
  {Hershfield}},\ }\href@noop {} {\bibfield  {journal} {\bibinfo  {journal}
  {Phys. Rev. Lett.}\ }\textbf {\bibinfo {volume} {70}},\ \bibinfo {pages}
  {2134} (\bibinfo {year} {1993})}\BibitemShut {NoStop}%
\bibitem [{\citenamefont {Han}(2006)}]{Han_06PRB}%
  \BibitemOpen
  \bibfield  {author} {\bibinfo {author} {\bibfnamefont {J.~E.}\ \bibnamefont
  {Han}},\ }\href@noop {} {\bibfield  {journal} {\bibinfo  {journal} {Phys.
  Rev. B}\ }\textbf {\bibinfo {volume} {73}},\ \bibinfo {pages} {125319}
  (\bibinfo {year} {2006})}\BibitemShut {NoStop}%
\bibitem [{\citenamefont {Anders}(2008)}]{Anders_08PRL}%
  \BibitemOpen
  \bibfield  {author} {\bibinfo {author} {\bibfnamefont {F.~B.}\ \bibnamefont
  {Anders}},\ }\href@noop {} {\bibfield  {journal} {\bibinfo  {journal} {Phys.
  Rev. Lett.}\ }\textbf {\bibinfo {volume} {101}},\ \bibinfo {pages} {066804}
  (\bibinfo {year} {2008})}\BibitemShut {NoStop}%
\bibitem [{\citenamefont {Schiller}\ and\ \citenamefont
  {Hershfield}(1995)}]{Schiller-Hershfield_95PRB}%
  \BibitemOpen
  \bibfield  {author} {\bibinfo {author} {\bibfnamefont {A.}~\bibnamefont
  {Schiller}}\ and\ \bibinfo {author} {\bibfnamefont {S.}~\bibnamefont
  {Hershfield}},\ }\href@noop {} {\bibfield  {journal} {\bibinfo  {journal}
  {Phys. Rev. B}\ }\textbf {\bibinfo {volume} {51}},\ \bibinfo {pages} {12896}
  (\bibinfo {year} {1995})}\BibitemShut {NoStop}%
\bibitem [{\citenamefont {Schiller}\ and\ \citenamefont
  {Hershfield}(1998)}]{Schiller-Hershfield_98PRB}%
  \BibitemOpen
  \bibfield  {author} {\bibinfo {author} {\bibfnamefont {A.}~\bibnamefont
  {Schiller}}\ and\ \bibinfo {author} {\bibfnamefont {S.}~\bibnamefont
  {Hershfield}},\ }\href@noop {} {\bibfield  {journal} {\bibinfo  {journal}
  {Phys. Rev. B}\ }\textbf {\bibinfo {volume} {58}},\ \bibinfo {pages} {14978}
  (\bibinfo {year} {1998})}\BibitemShut {NoStop}%
\bibitem [{\citenamefont {Shen}\ and\ \citenamefont
  {Fan}(2007)}]{Shen-Fan_07PRL}%
  \BibitemOpen
  \bibfield  {author} {\bibinfo {author} {\bibfnamefont {J.~T.}\ \bibnamefont
  {Shen}}\ and\ \bibinfo {author} {\bibfnamefont {S.}~\bibnamefont {Fan}},\
  }\href@noop {} {\bibfield  {journal} {\bibinfo  {journal} {Phys.\ Rev.\
  Lett.}\ }\textbf {\bibinfo {volume} {98}},\ \bibinfo {pages} {153003}
  (\bibinfo {year} {2007})}\BibitemShut {NoStop}%
\bibitem [{\citenamefont {Ordonez}\ \emph {et~al.}()\citenamefont {Ordonez},
  \citenamefont {Nishino},\ and\ \citenamefont
  {Hatano}}]{Ordonez-Nishino-Hatano_14preprint}%
  \BibitemOpen
  \bibfield  {author} {\bibinfo {author} {\bibfnamefont {G.}~\bibnamefont
  {Ordonez}}, \bibinfo {author} {\bibfnamefont {A.}~\bibnamefont {Nishino}}, \
  and\ \bibinfo {author} {\bibfnamefont {N.}~\bibnamefont {Hatano}},\
  }\href@noop {} {\bibinfo  {journal} {unpublished}}\BibitemShut {NoStop}%
\end{thebibliography}

%

\end{document}